\documentclass[prd,onecolumn,amsmath,amssymb,floatfix,nofootinbib]{revtex4}

\usepackage{amssymb}
\usepackage{amsmath}

\usepackage{graphicx}
\usepackage{graphics}
\usepackage{dcolumn}
\usepackage{color}
\usepackage{rotate}
\usepackage{fancyhdr}

\def\be{\begin{equation}}
\def\ee{\end{equation}}
\def\bea{\begin{eqnarray}}
\def\eea{\end{eqnarray}}
\def\ba{\begin{eqnarray}}
\def\ea{\end{eqnarray}}
\def\nn{\nonumber}
\def\vs{\nonumber\\ }

\def\VEV#1{\left\langle #1 \right\rangle}

\newcommand\bfepsilon{{\mathbf{\hat \varepsilon}}}
\newcommand{\hatk}{{\mathbf{\hat k}}}

\newcommand{\hatn}{{\mathbf{\hat n}}}

\newcommand{\hatx}{{\mathbf{\hat x}}}

\newcommand{\bfx}{\mathbf{x}}
\newcommand{\bfk}{\mathbf{k}}

\definecolor{royalblue}{rgb}{.25,.41,.88}

\definecolor{darkred}{rgb}{.743,0,0}

\def\bm#1{\mathbf{#1}}
\def\nhat{\hat{\bm{n}}}
\def\khat{\hat{\bm{k}}}

\def\CG#1#2#3#4#5#6{ \VEV{#1#2#3#4|#5#6}}

\def\wigner#1#2#3#4#5#6{ \left( \begin{array}{ccc} #1 & #3 & #5
\\ #2 & #4 & #6 \\ \end{array} \right)}

\begin{document}

\title{Wigner-Eckart theorem in cosmology: Bispectra for total-angular-momentum waves}

\author{Liang Dai, Donghui Jeong, and Marc Kamionkowski}
\affiliation{Department of Physics and Astronomy, Johns
     Hopkins University, 3400 N.\ Charles St., Baltimore, MD 21218}

\date{\today}

\begin{abstract}
Total-angular-momentum (TAM) waves provide a set of basis functions
for scalar, vector, and tensor fields that can be used in place
of plane waves and that reflect the rotational symmetry of the
spherical sky.  Here we discuss three-point correlation
functions, or bispectra in harmonic space, for scalar, vector,
and tensor fields in terms of TAM waves.  The Wigner-Eckart
theorem dictates that the expectation value, assuming statistical
isotropy, of the product of three TAM waves is the product of a
Clebsch-Gordan coefficient (or Wigner-3j symbol) times a
function only of the total-angular-momentum quantum numbers.
Here we show how this works, and we provide explicit expressions
relating the bispectra for TAM waves in terms of the more
commonly used Fourier-space bispectra.  This formalism will be
useful to simplify calculations of projections of
three-dimensional bispectra onto the spherical sky.
\end{abstract}

\maketitle

\section{Introduction}

Cosmological measurements have over the past several decades
made the notion of a period of
inflationary expansion in the early Universe particularly
appealing.  The principal aim of early-Universe cosmology has
thus become the elucidation of the new physics responsible for
inflation.  While the simplest single-field slow-roll (SFSR)
inflation models predict primordial perturbations to be very
nearly Gaussian \cite{inflation}, they do require some
nonvanishing departures from Gaussianity \cite{localmodel}.
Moreover, just about any extension of SFSR models, or embedding
of toy SFSR models into more realistic models
\cite{larger,curvaton,Dvali:1998pa}, or alternatives or
additions to inflation \cite{topdefects} lead to larger
departures from Gaussianity.  There has thus been ever-growing
attention focused on the search for non-Gaussianity.

The vast majority of the literature on departures from
Gaussianity focuses on the bispectrum (the three-point correlation
function in Fourier space) for the primordial curvature, a
scalar quantity \cite{NGReview}.  However, inflation also predicts primordial
gravitational waves (tensor metric perturbations)
\cite{Rubakov:1982df}, and some inflationary models involve the
introduction of vector fields \cite{vectors}. 
There are also discussions of primordial magnetic fields
\cite{Turner:1987bw}.  Just as there may arise non-Gaussianity
in the primordial scalar perturbation, there may also be
non-Gaussianity in these vector fields, in magnetic fields
\cite{Caldwell:2011ra,Barnaby:2012tk}, and in the gravitational-wave background
\cite{Maldacena:2002vr,Seery:2008ax,Jeong:2012df}.  Additional
work on non-Gaussianities with vector and scalar fields can be
found in Refs.~\cite{additional}.

Non-Gaussianity in the primordial scalar perturbation is most
commonly parametrized in terms of a bispectrum, the expectation
value for the product of three Fourier modes of wavevectors
$\bfk_1$, $\bfk_2$, and $\bfk_3$, of the fields under
consideration.  If the fields under consideration include vector
or tensor fields, then the bispectrum will also depend on some
contractions of the polarization vectors/tensors for these
fields.

Many cosmological observations, however, are performed on a
spherical sky, and many observables (e.g., CMB
temperature/polarization) depend only on an angular position on
the sky.  While others may depend on three-dimensional
positions, astronomical measurements discriminate
between the two angular coordinates and a radial coordinate.
Comparison of measured quantities with theoretical Fourier-space
bispectra therefore necessarily require projection of the
three-dimensional Fourier-space bispectra onto the
two-dimensional spherical sky.  The bispectra projected onto the
spherical sky depend on the three multipoles $l_1$, $l_2$, and
$l_3$, associated with the three spherical-harmonic coefficients
being correlated.  Such projections necessarily
involve the complications associated with spherical harmonics,
Clebsch-Gordan coefficients, and Wigner-3$j$ symbols, and
sometimes even Wigner-6$j$ and Wigner-9$j$ symbols.

In a recent paper \cite{Dai:2012}, we developed a
total-angular-momentum (TAM) formalism to provide sets of basis
functions for scalar, vector, and tensor fields
in three spatial dimensions that reflect the rotational
symmetry of the Universe about any given point (taken to be our
location).  The purpose of these TAM waves is to
incorporate the rotational symmetry of the sky from the start,
rather than start with plane waves which are later projected
onto the sky.  These TAM waves generalize the more familiar
tensor spherical harmonics
\cite{NewmanPenrose,Kamionkowski:1996ks,Zaldarriaga:1997pol},
which provide a basis for tensor functions on the
two-sphere, to a basis for functions in three-dimensional
Euclidean space.  The TAM waves of Ref.~\cite{Dai:2012} are also
provided for scalar and vector fields, and for all five
components of a traceless tensor, not just the two transverse
components.  

Each TAM wave is labeled by a
wavenumber $k$, total-angular-momentum quantum numbers $J$, and
azimuthal quantum number $M$.  There are
three sets of TAM waves for each $kJM$ for vector fields (to
reflect the three components of a vector field) and five sets of
TAM waves for each $kJM$ for a symmetric traceless rank-2 tensor.  We
provided in Ref.~\cite{Dai:2012} several different sets of
bases, for a given $JM$, for these three vector TAM waves and
five tensor TAM waves.  In the orbital-angular-momentum (OAM)
basis, the three vector basis functions for a given $JM$ have
orbital angular momentum $l=J-1,J,J+1$, and the five tensor basis
functions have $l=J-2,J-1,J,J+1,J+2$.  In the second basis,
the three vector basis functions for a given $JM$ include a
longitudinal ($L$) mode and two transverse modes ($E$ and $B$)
of opposite parity.  The five tensor functions in this basis
include a longitudinal ($L$) mode, two vector modes ($VE$ and
$VB$) of opposite parity, and two transverse-tensor modes ($TE$
and $TB$) of opposite parity.  The third basis represents vector
fields of a given $JM$ in terms of states of helicity
$\lambda=0,\pm1$ and tensor fields in terms of modes of helicity
$\lambda=0,\pm1,\pm2$.   The most general scalar, vector, or tensor
field can then be written in terms of these TAM waves, rather
than Fourier modes.  Spherical-sky observables are then obtained
more naturally from these TAM waves than from Fourier waves.

In this paper we calculate the bispectra for TAM waves in terms
of the more commonly used Fourier-space bispectra. As we will
see, the Wigner-Eckart theorem guarantees that the expectation value of the
product of three TAM-wave coefficients depends on $M_1$, $M_2$,
and $M_3$ only through the Clebsch-Gordan coefficient (or
equivalently Wigner-3$j$ symbol or Gaunt integral);  there may
then be some dependence on $J_1$, $J_2$, and $J_3$ in a
prefactor in addition to that in the Clebsch-Gordan coefficient.
The specific form for the prefactor will 
depend on whether the bispectrum is for scalar, vector, and/or
tensor fields as well as on the tensorial nature of the
bispectrum (i.e., how the indices on the polarization
vector/tensors are contracted).  The principal results of this
paper are thus the specific forms for these prefactors for a
variety of bispectra.  These results will facilitate the
calculation of observables, such as angular bispectra, for
theories that involve such correlations, particularly those that
involve vector and/or tensor fields.

We begin in Section \ref{sec:TAMreview} with a brief review of
the total-angular-momentum wave (TAM) formalism, and we use the
notation and conventions of Ref.~\cite{Dai:2012} throughout.
Section \ref{sec:ThreeS} presents the bispectra for three scalar
TAM waves.  Section~\ref{sec:twoVoneS} follows to calculate
bispectra involving two transverse-vector TAM waves, in the
$E/B$ basis, and one scalar TAM wave.
Section~\ref{sec:twoToneS} presents bispectra of two
transverse-traceless tensor TAM waves plus one scalar TAM
wave.  Section~\ref{sec:oneTtwoS} then
deals with the correlation of one symmetric traceless tensor
TAM wave with two scalar TAM waves, where the traceless tensor
can be a transverse-traceless, vectorial, or longitudinal.
The same results apply also to correlation of
one traceless tensor TAM wave and two longitudinal
vector TAM waves.  
Section~\ref{sec:conclusion} provides concluding remarks and
provides a list of the equation numbers for the central results
that relate the Fourier-space and TAM-wave three-point functions.
Appendix~\ref{app:3Ylms_LEB} presents helicity-basis overlap
integrals that are used earlier in the paper. We discuss a
semi-classical picture to understand the structure of the
TAM-wave bispectra in Appendix~\ref{app:semi_classical}.

\section{Review of Total Angular Momentum Waves}
\label{sec:TAMreview}

Here we very briefly re-introduce total-angular-momentum (TAM)
waves for scalar, vector, and tensor fields.  We follow
throughout the convention and notation of Ref.~\cite{Dai:2012} and
refer the reader there for more details.

TAM waves are eigenfunctions of eigenvalue $-k^2$ of the
Laplacian that are also eigenstates of quantum numbers $J$ and
$M$ of total angular momentum squared and its third component.  The
scalar TAM waves are denoted by $\Psi^k_{(JM)}(\bfx)$.  
TAM vector waves of orbital angular momentum $l$ are specified by
$\Psi^{l,k}_{(JM)a}(\bfx)$, where $a=\{x,y,z\}$ is the vector index
and $l=J-1,J,J+1$.  TAM tensor waves of orbital angular momentum
$l$ are specified by $\Psi^{l,k}_{(JM)ab}(\bfx)$, where $a$ and
$b$ are the tensor indices and $l=J-2,J-1,J,J+1,J+2$.  The
scalar, vector, and tensor TAM waves are distinguished by the
number of indices.

These TAM waves satisfy the completeness relations,
\ba
     \sum_{JM} \int \frac{k^2\, dk}{(2\pi)^3} 
     \left[ 4 \pi i^J
     \Psi^k_{(JM)}(\bfx) \right]^*  
\left[ 4 \pi i^J
     \Psi^k_{(JM)}(\bfx') \right] &=& \delta_D(\bfx-\bfx'), \\
     \sum_{JMl} \int \frac{k^2\, dk}{(2\pi)^3} \left[ 4 \pi i^l
     \Psi^{l,k}_{(JM)a}(\bfx) \right]^*  \left[ 4 \pi i^l
     \Psi^{l,k}_{(JM)}{}^a(\bfx') \right] &=& \delta_D(\bfx-\bfx'), \\
     \sum_{JMl} \int \frac{k^2\, dk}{(2\pi)^3} \left[ 4 \pi i^l
     \Psi^{l,k}_{(JM)ab}(\bfx) \right]^*  \left[ 4 \pi i^l
     \Psi^{l,k}_{(JM)}{}^{ab}(\bfx') \right] &=&
     \delta_D(\bfx-\bfx').
\ea
These relations imply that an arbitrary scalar, vector, or
tensor field can be expanded, respectively,
\ba
     \phi(\bfx) &=&      \sum_{JM} \int \frac{k^2\, dk}{(2\pi)^3}
     \phi_{(JM)}^k \, 4\pi i^J \Psi^k_{(JM)}(\bfx), \\
     V_a(\bfx) &=&      \sum_{JMl} \int \frac{k^2\, dk}{(2\pi)^3}
     V_{(JM)}^{l,k} \, 4\pi i^l \Psi^{l,k}_{(JM)a}(\bfx), \\
     T_{ab}(\bfx) &=&      \sum_{JMl} \int \frac{k^2\, dk}{(2\pi)^3}
     T_{(JM)}^{l,k} \, 4\pi i^l \Psi^{l,k}_{(JM)ab}(\bfx),
\ea
in terms of TAM waves and expansion coefficients,
\ba
    \phi_{(JM)}^k & = & \int \, d^3x\, \phi(\bfx) 
     \left[ 4 \pi i^J \Psi^k_{(JM)}(\bfx) \right]^*,  \\
    V^{l,k}_{(JM)} & = & \int \, d^3x\, V^a(\bfx) 
     \left[ 4 \pi i^l \Psi^{l,k}_{(JM)a}(\bfx) \right]^*,  \\
    T^{l,k}_{(JM)} & = & \int \, d^3x\, T^{ab}(\bfx) 
     \left[ 4 \pi i^l \Psi^{l,k}_{(JM)a}(\bfx) \right]^*.
\ea
By using the plane-wave expansions,
\ba
     e^{i \bfk \cdot \bfx} &=& \sum_{JM} 4 \pi i^J \left[
     Y_{(JM)}^*(\hatk) \right]
     \Psi^k_{(JM)}(\bfx), \label{eqn:scalarplanewave} \\
     \hat\varepsilon_a(\hatk) e^{i \bfk \cdot \bfx} &=&
     \sum_{JMl} 4 \pi i^l \left[ \hat\varepsilon^b(\hatk)
     Y^{l\,*}_{(JM)b}(\hatk) \right] \Psi^{l,k}_{(JM)a}(\bfx),\\
     \hat\varepsilon_{ab}(\hatk) e^{i \bfk \cdot \bfx} &=&
     \sum_{JMl} 4 \pi i^l \left[ \hat\varepsilon^{cd}(\hatk)
     Y^{l\,*}_{(JM)cd}(\hatk) \right] \Psi^{l,k}_{(JM)ab}(\bfx),
\ea
and orthogonality of the TAM waves,
\ba
     \int\, d^3x\, \left[ 4\pi i^{J'} \Psi^{k'}_{(J'M')}(\bfx)
     \right]^* 
     \left[ 4\pi i^{J} \Psi^{k}_{(JM)}(\bfx) \right] &=&
     \delta_{JJ'} \delta_{MM'} \frac{(2\pi)^3}{k^2}
     \delta_D(k-k'), \\
     \int\, d^3x\, \left[ 4\pi i^{l'} \Psi^{l',k'\,\,a}_{(J'M')}(\bfx)
     \right]^* 
     \left[ 4\pi i^{l} \Psi^{l,k}_{(JM)a}(\bfx) \right] &=&
     \delta_{ll'} \delta_{JJ'} \delta_{MM'} \frac{(2\pi)^3}{k^2}
     \delta_D(k-k'), \\
     \int\, d^3x\, \left[ 4\pi i^{l'} \Psi^{l',k'\,\,ab}_{(J'M')}(\bfx)
     \right]^* 
     \left[ 4\pi i^{l} \Psi^{l,k}_{(JM)ab}(\bfx) \right] &=&
     \delta_{ll'} \delta_{JJ'} \delta_{MM'} \frac{(2\pi)^3}{k^2}
     \delta_D(k-k'),
\ea
we can also write the expansion coefficients for scalar, vector
and tensor waves as
\ba
     \phi^k_{(JM)} &=& 
     \int d^2\hatk \, \tilde \phi(\bfk)
     Y_{(JM)}^*(\hatk) \label{eqn:scalarspherical}, \\
     V^{l,k}_{(JM)} &=& 
       \int d^2\hatk\, \tilde
     V^a(\bfk) Y^{l\,\,*}_{(JM)a}(\hatk)\label{eqn:vectorspherical}, \\
     T^{l,k}_{(JM)} &=& 
	\int \, d^2\hatk\, \tilde T^{ab}(\bfk)
     Y^{l\,\,*}_{(JM)ab}(\hatk). \label{eqn:tensorspherical} 
\ea

Similarly, if we choose to decompose the vector field into
$L/E/B$ waves and the tensor fields into $L/VE/VB/TE/TB$ waves,
which we label by $\alpha$, then we also have
\ba
     V^{\alpha,k}_{(JM)} &=& \int \, d^3x\, V^a(\bfx)\left[4 \pi i^J
     \Psi^{\alpha,k}_{(JM)a}(\bfx) \right]^* = \int\, d^2
     \hatk\, \tilde V^a(\bfk) Y^{\alpha\,\,*}_{(JM)a}(\hatk) ,
		\label{eqn:vectorLEBspherical}
		\\
     T^{\alpha,k}_{(JM)} &=& \int \, d^3x\, T^{ab}(\bfx) \left[4 \pi i^J
     \Psi^{\alpha,k}_{(JM)a}(\bfx) \right]^* = \int\, d^2
     \hatk\, \tilde T^{ab}(\bfk) Y^{\alpha\,\,*}_{(JM)ab}(\hatk).
		\label{eqn:tensorLEBspherical}
\ea

Finally, we can write the $L/E/B$ vector TAM waves in terms of the
$L/E/B$ vector spherical harmonics as
\begin{eqnarray}
     \Psi^{B,k}_{(JM)a}(\bfx) &=& j_J(kr) Y^B_{(JM)a}(\hatn), \nn \\
     \Psi^{E,k}_{(JM)a}(\bfx) &=& -i\left[ j_J'(kr)
     +\frac{j_J(kr)}{kr} \right] Y^E_{(JM)a}(\hatn) -
     i\sqrt{J(J+1)} \frac{j_J(kr)}{kr} Y^L_{(JM)a}(\hatn), \nn
     \\
     \Psi^{L,k}_{(JM)a}(\bfx) &=& -i\sqrt{J(J+1)}
     \frac{j_J(kr)}{kr} Y^E_{(JM)a}(\hatn) - ij_J'(kr)
     Y^L_{(JM)a}(\hatn),
\label{eqn:vectorprojections}
\end{eqnarray}
and the $L/VE/VB/TE/TB$ tensor TAM waves in terms of the 
$L/VE/VB/TE/TB$ tensor spherical harmonics as 
\begin{align}
     \Psi_{(JM)ab}^{L,k}(\bfx) = &
     -\frac{1}{2}\left(j_{J}(kr)+3j_{J}''(kr)\right)Y_{(JM)ab}^{L}
     (\hatn) -\sqrt{3J(J+1)}f_{J}(kr)Y_{(JM)ab}^{VE}(\hatn)\nn\\ 
   & - \frac{1}{2} \sqrt{\frac{3 (J+2)!}{(J-2)!}}
   \frac{j_{J}(kr)}{(kr)^2}Y_{(JM)ab}^{TE}(\hatn),\nn\\
   \Psi_{(JM)ab}^{VE,k}(\bfx)=&-\sqrt{3J(J+1)}f_{J}(kr)Y_{(JM)ab}^{L}(\hatn)-\left(j_{J}(kr)+2j_{J}''(kr)+2f_{J}(kr)\right)Y_{(JM)ab}^{VE}(\hatn)\nn\\
&-\sqrt{\left(J-1\right)\left(J+2\right)}\left(f_{J}(kr)+2\frac{j_{J}(kr)}{(kr)^2}\right)Y_{(JM)ab}^{TE}(\hatn),\nn\\
     \Psi_{(JM)ab}^{TE,k}(\bfx)=&-\frac{1}{2}
     \sqrt{\frac{3(J+2)!}{(J-2)!}}
     \frac{j_{J}(kr)}{(kr)^2}Y_{(JM)ab}^{L}(\hatn)-\sqrt{\left(J-1\right)\left(J+2\right)}\left(f_{J}(kr)+2\frac{j_{J}(kr)}{(kr)^2}\right)Y_{(JM)ab}^{VE}(\hatn)\nn\\
&-\frac12\left(-j_{J}(kr)+j_{J}''(kr)+4f_{J}(kr)+6\frac{j_{J}(kr)}{(kr)^2}\right)Y_{(JM)ab}^{TE}(\hatn),\nn\\
\Psi_{(JM)ab}^{VB,k}(\bfx)=&-i\left(j_{J}'(kr)-\frac{j_{J}(kr)}{kr}\right)Y_{(JM)ab}^{VB}(\hatn)-i\sqrt{\left(J-1\right)\left(J+2\right)}\frac{j_{J}(kr)}{kr}Y_{(JM)ab}^{TB}(\hatn),\nn\\
\Psi_{(JM)ab}^{TB,k}(\bfx)=&-i\sqrt{\left(J-1\right)\left(J+2\right)}\frac{j_{J}(kr)}{kr}Y_{(JM)ab}^{VB}(\hatn)-i\left(j_{J}'(kr)+2\frac{j_{J}(kr)}{kr}\right)Y_{(JM)ab}^{TB}(\hatn),
\label{eqn:tensorprojections}
\end{align}
where $f_J(x)=(j_J(x)/x)'$. For later use, we define additional radial functions for vector
TAM waves with $\alpha,\beta=E,L$ by
\be
     \Psi_{(JM) a}^{\alpha, k}(\bfx) 
     \equiv 
     -i \sum_{\beta} j_{J,v}^{(\alpha,\beta)}(kr)Y_{(JM) a}^{\beta}(\nhat),
\label{eqn:def_jJvec}
\ee
and for tensor TAM waves with $\alpha,\beta = L,VE,TE$ by
\be
\Psi_{(JM) ab}^{\alpha, k}(\bfx) 
\equiv 
- \sum_{\beta} j_{J,t}^{(\alpha,\beta)}(kr)Y_{(JM) ab}^{\beta}(\nhat),
\label{eqn:def_jJten1}
\ee
and for tensor TAM waves with $\alpha,\beta = TB,VB$ by
\be
\Psi_{(JM) ab}^{\alpha, k}(\bfx) 
\equiv 
- i\sum_{\beta} j_{J,t}^{(\alpha,\beta)}(kr)Y_{(JM) ab}^{\beta}(\nhat).
\label{eqn:def_jJten2}
\ee
The precise forms for $j_{J,v}^{(\alpha,\beta)}$ can be read off
Eq.~(\ref{eqn:vectorprojections}) and for
$j_{J,t}^{(\alpha,\beta)}$ from Eq.~(\ref{eqn:tensorprojections}).
Some of these radial profiles have appeared in previous full-sky
studies in cosmology. For example in the line-of-sight approach
to CMB polarization, the full-sky EE and BB power spectra,
Eq.(29) of Ref.~\cite{Zaldarriaga:1997pol}), contain radial
kernels identical to the functions $j^{(TE,TE)}_{J,t}(x)$ and
$j^{(TB,TB)}_{J,t}(x)$ defined here.

\subsection{From TAM-bispectrum to angular bispectrum}

Throughout this paper, we will calculate TAM bispectra of the form,
\be
\left\langle
X_{J_1M_1}^{\alpha_1,\, k_1}
Y_{J_2M_2}^{\alpha_2,\, k_2}
Z_{J_3M_3}^{\alpha_3,\, k_3}
\right\rangle,
\ee
where $X$, $Y$, and $Z$ represent generic scalar, vector and
tensor fields, and $\alpha_i$ represents the particular mode
of interest ($L/E/B$ for vector fields and $L/VE/VB/TE/TB$ for
tensor fields).
The TAM-wave bispectra above are related to the angular
bispectra for various observables as follows.  Suppose we
observe some projection of a three-dimensional field $X(\bfx)$.
The spherical-harmonic coefficients for the projection can be
written in terms of the Fourier components $\tilde X(\bfk)$ of
the field as
\be
     a_{lm}^{\alpha\,X} = 
     \int \frac{d^3k}{(2\pi)^3} g_{X}(k)
     \tilde{X}(\bm{k}) \mathcal{Y}_{lm}^{\alpha\,*}(\khat)
     =
     \int  \frac{k^2 dk}{(2\pi)^3} g_X(k) 
     X_{(lm)}^{\alpha,\,k},
\ee
in terms of a transfer function $g_X(k)$ which, assuming
statistical isotropy, is a function only of the wavevector
magnitude $k$.  Here we suppress tensor indices, as the end
result (as will be spelled out more clearly in the rest of the
paper) is the same for scalar, vector and tensor fields, and
$\mathcal{Y}_{lm}^\alpha$ generically represents the
corresponding $\alpha$ modes of scalar, vector, and tensor
spherical harmonics.
The observable angular bispectrum is then related to the TAM
bispectrum as 
\be
\left\langle
a_{J_1M_1}^{\alpha_1,X}
a_{J_2M_2}^{\alpha_2,Y}
a_{J_3M_3}^{\alpha_3,Z}
\right\rangle
=
\int  \frac{k_1^2 dk_1}{(2\pi)^3} g_X(k_1) 
\int  \frac{k_2^2 dk_2}{(2\pi)^3} g_Y(k_2) 
\int  \frac{k_3^2 dk_3}{(2\pi)^3} g_Z(k_3) 
\left\langle
X_{J_1M_1}^{\alpha_1,\, k_1}
Y_{J_2M_2}^{\alpha_2,\, k_2}
Z_{J_3M_3}^{\alpha_3,\, k_3}
\right\rangle.
\ee

For example, if we take $X(\bfx)$ to be Bardeen's curvature
perturbation $\Phi(\bfx)$, and the $a_{lm}^X$ to be
CMB-temperature spherical-harmonic coefficients, then the
appropriate transfer function is
\be
g_{T}(k)
=
4\pi i^l g_{Tl}(k),
\ee
in terms of the radiation transfer function $g_{Tl}(k)$
\cite{Komatsu:2001rj}.  If the $a_{lm}^X$ are taken to be the
density of galaxies on the sky, then the transfer function is
\be
g_g(k)
= 
4\pi i^l b_g
\left(\frac23 \frac{k^2 T(k)}{H_0^2\Omega_m}\right)
\int dz W_g(z)
D(z) j_l[kd_A(z)],
\ee
with galaxy bias $b_g$, matter transfer function $T(k)$,,
linear growth factor $D(z)$, comoving angular=diameter distance $d_A(z)$,
and redshift distribution $W_g(z)$ of galaxies normalized to 
$\int dz W_g(z)=1$. Here, $H_0$ and $\Omega_m$ are, respectively,
the Hubble and matter-density parameters at present.

\section{Scalar bispectrum}
\label{sec:ThreeS}

Before turning to vector and tensor fields, for which the TAM
formalism provides the most substantial advantage, we begin by
way of introduction with scalar bispectra in the TAM formalism.

The bispectrum $B_{sss}(k_1,k_2,k_3)$ for the scalar field is usually
defined in terms of Fourier modes as
\be
     \VEV{ \tilde \phi(\bfk_1) \tilde \phi(\bfk_2) \tilde
     \phi(\bfk_3) } = (2\pi)^3
     \delta_D(\bfk_1 + \bfk_2 + \bfk_3) B_{sss}(k_1,k_2,k_3),
\label{eqn:Fourierbispectrum}
\ee
where the Dirac delta function arises as a consequence of
statistical homogeneity, and the dependence of the
bispectrum only on the magnitudes $k_1$, $k_2$, and $k_3$ as a
consequence of statistical isotropy.  The
subscript `$sss$' is used to denote the bispectrum for three
scalar fields, to distinguish it from the bispectra to be discussed
below that involve vector and/or tensor fields.

If the scalar bispectrum is defined as in
Eq.~(\ref{eqn:Fourierbispectrum}), then the expectation value
for the product of three TAM-wave coefficients are given by
\be
     \VEV{ \phi_{(J_1M_1)}^{k_1} \phi_{(J_2 M_2)}^{k_2} \phi_{(J_3
     M_3)}^{k_3} } = \int d^2\hatk_1 \int d^2\hatk_2 \int
     d^2\hatk_3 \VEV{ \tilde \phi(\bfk_1) \tilde \phi(\bfk_2)
     \tilde \phi(\bfk_3) } Y^*_{(J_1 M_1)}(\hatk_1) Y^*_{(J_2 M_2)}(\hatk_2)
     Y^*_{(J_3 M_3)}(\hatk_3).
\ee
We then substitute Eq.~(\ref{eqn:Fourierbispectrum}) and expand
the Dirac delta function, using the plane-wave expansion,
Eq.~(\ref{eqn:scalarplanewave}), as
\ba
     (2\pi)^3 \delta_D(\bfk_1 + \bfk_2 + \bfk_3) &=& \int d^3x\,
     e^{-i \bfk_1 \cdot \bfx} e^{-i \bfk_2 \cdot \bfx} e^{-i
     \bfk_3\cdot \bfx}
	\nn \\
		&=&  \int d^3x \sum_{l_im_i} (4\pi)^3 (-i)^{l_1+l_2+l_3}
     j_{l_1}(k_1 r) j_{l_2}(k_2 r) j_{l_3}(k_3 r) \nn\\
     & & \qquad \times Y_{(l_1 m_1)}(\hatk_1) Y_{(l_1
     m_1)}^*(\hatn) Y_{(l_2 m_2)}(\hatk_2) Y_{(l_2 m_2)}^*(\hatn)
     Y_{(l_3 m_3)}(\hatk_3) Y_{(l_3 m_3)}^*(\hatn),
\label{eqn:Diracexpansion}
\ea
to eliminate the $\hatk_i$ integrals.  We can then write the
TAM-wave bispectrum as
\be
     \VEV{ \phi_{(J_1M_1)}^{k_1} \phi_{(J_2 M_2)}^{k_2} \phi_{(J_3
     M_3)}^{k_3} } = (4\pi)^3 (-i)^{J_1+J_2+J_3} B_{sss}(k_1,k_2,k_3)
     \int d^3x\, \Psi^{k_1\,\,*}_{(J_1M_1)}(\bfx)
     \Psi^{k_2\,\,*}_{(J_2M_2)}(\bfx)
     \Psi^{k_3\,\,*}_{(J_3M_3)}(\bfx),  
\ee
in terms of an overlap of three TAM waves.  Below we will see
that the TAM-wave three-point functions will always depend, even
with vector and tensor fields, on overlaps of three TAM waves.
Since the scalar TAM
waves are $\Psi^{k}_{JM}(\bfx) = j_J(kr) Y_{(JM)}(\hatx)$, the
overlap is simply, 
\begin{equation}
     \int \,d^3x\, \Psi^{k_1}_{J_1M_1}(\bfx) \Psi^{k_2}_{J_2
     M_2}(\bfx)  \Psi^{k_3}_{J_3 M_3}(\bfx) = {\cal
     G}^{J_1J_2J_3}_{M_1 M_2 M_3} J_{J_1 J_2 J_3}(k_1,k_2,k_3),
\end{equation}
where
\begin{equation}
     {\cal G}_{m_1m_2 m_3}^{l_1 l_2 l_3} = \int \, d^2 \hatn
     Y_{(l_1 m_1)}(\hatn) Y_{(l_2
     m_2)}(\hatn) Y_{(l_3 m_3)}(\hatn) = \sqrt{ \frac{
     (2l_1+1) (2l_2+1) (2l_3+1) }{4\pi}}
     \wigner{l_1}{m_1}{l_2}{m_2}{l_3}{m_3}
     \wigner{l_1}{0}{l_2}{0}{l_3}{0},
\end{equation}
is the Gaunt integral, and
\begin{equation}
     J_{l_1 l_2 l_3}(k_1,k_2,k_3) = \int_0^\infty\, r^2\, dr\, 
     j_{l_1}(k_1 r) j_{l_2}(k_2 r) j_{l_3}(k_3 r).
\label{eqn:radialintegral}
\end{equation}
Therefore,
\be
     \VEV{ \phi_{(J_1M_1)}^{k_1} \phi_{(J_2 M_2)}^{k_2} \phi_{(J_3
     M_3)}^{k_3} } = (4\pi)^3 (-i)^{J_1+J_2+J_3} {\cal
     G}^{J_1J_2J_3}_{M_1 M_2 M_3} 
		J_{J_1 J_2 J_3}(k_1,k_2,k_3)
     B_{sss}(k_1,k_2,k_3).
\label{eqn:scalarfinal}
\ee
We have thus evaluated the TAM-wave three-point correlation
function in terms of the overlap of three TAM waves.  We then
arrive at our final result, Eq.~(\ref{eqn:scalarfinal}), which
provides a simple expression for the TAM-wave three-point
function in terms of the Fourier-space bispectrum, by evaluating
that overlap.  The factors connecting the Fourier-space and
TAM-wave three-point functions are the easily evaluated and
familiar Gaunt integral, and the radial integral,
Eq.~(\ref{eqn:radialintegral}).  The Gaunt integral is
proportional to a Wigner-3j symbol and the Clebsch-Gordan
coefficient.  Its presence here is a result of the Wigner-Eckart
theorem, and we will see in the Sections below that the
expectation value for the product of any three TAM waves---not
just scalar TAM waves---will be proportional to a Clebsch-Gordan
coefficient times a function only of $J_1$, $J_2$, and $J_3$.
We choose to write the results in terms of the Gaunt factor,
rather than Clebsch-Gordan coefficients or Wigner-3j symbols
simply because it provides for more compact expressions.

The radial integral in Eq.~(\ref{eqn:scalarfinal}) can be
evaluated numerically using standard techniques
\cite{Mehrem:1991me,Wang,Toyoda:2010ty}, but often an end
result, which requires an additional integral over the $k_i$,
can be obtained more quickly by changing the orders of those
$k_i$ integrals and the radial integral in
Eq.~(\ref{eqn:radialintegral}).

Below we will generalize Eq.~(\ref{eqn:scalarfinal}) to
three-point functions involving vector and tensor fields as well.

\section{Two transverse vectors and a scalar}
\label{sec:twoVoneS}

We now proceed with the analogous calculation for the
three-point correlation function that involves two
transverse-vector fields and a scalar.  The transverse vector
field $V^a(\bfx)$ satisfies $\nabla_a V^a=0$, or in Fourier
space, $k_a \tilde V^a(\bfk)=0$.  Such a vector-vector-scalar
correlation arises, for example, in models in which a magnetic
field (the transverse-vector field) is produced during inflation
and thus correlated with the inflaton field (the scalar field)
\cite{Caldwell:2011ra}.

Our goal is to calculate the TAM-wave bispectra,
\be
     \VEV{V^{\alpha_1,k_1}_{(J_1M_1)} V^{\alpha_2,k_2}_{(J_2M_2)}
     \phi^{k_3}_{(J_3M_3)}},
\ee
for $\{\alpha_1,\alpha_2\}=\{E,B\}$ that correspond to a given
Fourier-space bispectrum.  

\subsection{Fourier-space bispectra}
\label{sec:fourierspace}

We begin by discussing the Fourier-space bispectrum.
Each Fourier component of the transverse-vector field can be
written
\be
     \tilde V^a(\bfk) = \sum_{\lambda=\pm}
     \hat\varepsilon_\lambda^a(\bfk) \tilde V_\lambda(\bfk),
\ee
in terms of two helicity-basis polarization vectors
$\bfepsilon_\lambda(\bfk)$ and helicity Fourier amplitudes
$\tilde V_\lambda(\bfk)$.  The helicity basis vectors can be
written as $\bfepsilon_\pm(\bfk) = \left[\bfepsilon_1(\bfk) \pm i \hatk
\times \bfepsilon_1(\bfk)\right]/\sqrt{2}$, where $\bfepsilon_1$
is any unit vector in the plane perpendicular to $\bfk$.
The most general Fourier-space vector-vector-scalar three-point
function can therefore be written,
\be
     \VEV{ \tilde V^a(\bfk_1) \tilde V^b(\bfk_2) \tilde \phi(\bfk_3)} =
     \sum_{\lambda_1 \lambda_2} \hat\varepsilon_{\lambda_1}^a(\bfk_1)
     \hat\varepsilon_{\lambda_2}^b(\bfk_2) \VEV{
     \tilde V_{\lambda_1}(\bfk_1) \tilde V_{\lambda_2}(\bfk_2)
     \tilde \phi(\bfk_3)}.
\label{eqn:first}
\ee
Given the orthonormality of the polarization vectors, this
relation can then be inverted to provide the three-point
correlation,
\be
   \VEV{
     \tilde V_{\lambda_1}(\bfk_1) \tilde V_{\lambda_2}(\bfk_2)
     \tilde \phi(\bfk_3)} = 
     \hat\varepsilon_{\lambda_1}^a(\bfk_1)
     \hat\varepsilon_{\lambda_2}^b(\bfk_2) \VEV{ V_a(\bfk_1)
     V_b(\bfk_2) \phi(\bfk_3)},
\label{eqn:crucial}
\ee
for the helicity amplitudes.

Now consider the tensor structure of the three-point
function on the right side of Eq.~(\ref{eqn:crucial}). By momentum conservation, it
most generally depends on $k_1^a$ and $k_2^a$. Among all structures one can possibly construct, $g^{ab}$, $k_1^a k_1^b$, 
$k_1^a k_2^b$, $k_2^a k_1^b$ and $k_2^a k_2^b$, terms proportional to $k_a^1$ or
$k_2^b$ have no contribution to Eq.~(\ref{eqn:crucial}) under
helicity basis, and hence to the bispectrum through
Eq.~(\ref{eqn:first}).\footnote{If parity violation is allowed,
there may be additional terms constructed
with the antisymmetric tensor $\epsilon_{abc}$, but we restrict
our attention to parity-preserving terms.} The
most general parity-conserving three-point function can then be written,
\be
     \VEV{ \tilde V^a(\bfk_1) \tilde V^b(\bfk_2) \tilde
     \phi(\bfk_3)} = (2\pi)^3 \delta_D(\bfk_1 +\bfk_2 + \bfk_3) \left[
     g^{ab}B_{vvs}^{(1)}(k_1,k_2,k_3) + k_2^a k_1^b
     B_{vvs}^{(2)}(k_1,k_2,k_3) \right],
\ee
in terms of two vector-vector-scalar bispectra
$B_{vvs}^{(1)}(k_1,k_2,k_3)$ and $B_{vvs}^{(2)}(k_1,k_2,k_3)$.
Two different bispectra arise because there
are two components for each polarization vector.

To demonstrate our formalism, we take the example of non-zero
 $B_{vvs}^{(1)}(k_1,k_2,k_3)$, which may arise
from a local three-point interaction of the form $\phi(\bfx) |{\bf
V}(\bfx)|^2$.  The other bispectrum $B_{vvs}^{(2)}(k_1,k_2,k_3)$, 
which can arise as the vector-vector-scalar correlation between the vector
eletromagnetic potential and the inflaton from a $\phi F_{\mu\nu}F^{\mu\nu}$
coupling \cite{Caldwell:2011ra}, is left to future work.

Therefore, the Fourier-space bispectrum we will consider here
will be of the form,
\be
     \VEV{\tilde V^a(\bfk_1) \tilde V^b(\bfk_2) \tilde \phi(\bfk_3)} = (2\pi)^3
     \delta_D(\bfk_1 +\bfk_2 + \bfk_3) g^{ab}
     B_{vvs}(k_1,k_2,k_3).
\label{eqn:vvs}
\ee

\subsection{TAM-wave bispectra}

We now calculate the three-point correlation function,
\be
     \VEV{V^{\alpha_1,k_1}_{(J_1M_1)} V^{\alpha_2,k_2}_{(J_2M_2)}
     \phi^{k_3}_{(J_3M_3)}} = \int d^2\hatk_1 d^2\hatk_2
     d^2\hatk_3\, Y^{\alpha_1\,\,*}_{(J_1 M_1)a}(\hatk_1)
     Y^{\alpha_2\,\,*}_{(J_2 M_2)b}(\hatk_2) Y^{\,\,*}_{(J_3
     M_3)}(\hatk_3) \VEV{\tilde V^a(\bfk_1) \tilde V^b(\bfk_2)
     \tilde \phi(\bfk_3)},
\label{eqn:overlaps}
\ee
for the $E/B$ TAM-wave coefficients, where
$\alpha_1,\alpha_2=\{E,B\}$, and we have used
Eq.~(\ref{eqn:vectorLEBspherical}).  The first step is to expand
the Dirac delta function in Eq.~(\ref{eqn:vvs}), and using
the identity,
\be
     \sum_{J'M'}
	  (-i)^{J'} j_{J'}(kr)
	  Y_{(J'M')}^*(\hatn) \int d^2 \hatk\, Y^{\alpha\,\,*}_{(JM)a}(\hatk)
     Y_{(J'M')}(\hatk) 
	  =(-i)^J \left[\Psi_{(JM) a}^{\alpha,k}(\bfx)\right]^*,
\ee
we find that 
\be
     \VEV{V^{\alpha_1,k_1}_{(J_1M_1)} V^{\alpha_2,k_2}_{(J_2M_2)}
     \phi^{k_3}_{(J_3M_3)}} = 
		(4\pi)^3 (-i)^{J_1+J_2+J_3}
     B_{vvs}(k_1,k_2,k_3) 
		\int d^3x\,
     \Psi^{\alpha_1,k_1\,\,*}_{(J_1M_1)}{}^a(\bfx) 
     \Psi^{\alpha_2,k_2\,\,*}_{(J_2 M_2)a}(\bfx)
     \Psi^{k_3\,\,*}_{(J_3M_3)}(\bfx).
\label{eqn:TAMbk_overlaps}
\ee
That is, the bispectrum of three TAM coefficients is proportional to the 
overlap of three TAM waves.

These overlap integrals can be evaluated with 
the projection of vector TAM waves in Eq.~(\ref{eqn:vectorprojections}).
The angular parts of the overlap integral
can be evaluated by writing the $L/E/B$ vector spherical
harmonics in terms of helicity spherical harmonics as
\be
     Y^{\pm1}_{(JM)a} = \left( Y^{E}_{(JM)a} \pm i Y^{B}_{(JM)a}
     \right)/\sqrt{2}, \qquad Y^L_{(JM)a}= Y^0_{(JM)a}.
\ee
We can then use the helicity-harmonic integrals in
Appendix~\ref{app:3Ylms_LEB_vector} to obtain the
spherical-harmonic overlap integrals,
\ba
     \int d^2\hatn Y_{(J_1M_1)}^{L,a}(\hatn)
     Y_{(J_2M_2)a}^{L}(\hatn) Y_{\left(J_3M_3\right)}
     (\hatn)  &=&  {\cal G}^{J_1 J_2 J_3}_{M_1 M_2 M_2},\nonumber \\
     \int d^2\hatn Y_{(J_1M_1)}^{B,a}(\hatn)
     Y_{(J_2M_2)a}^{B} (\hatn)
     Y_{\left(J_3M_3\right)}(\hatn) &=& \int
     d^2\hatn Y_{(J_1M_1)}^{E,a} (\hatn)
     Y_{(J_2M_2)a}^{E}(\hatn)
     Y_{\left(J_3M_3\right)}(\hatn)  \nonumber \\
     & = & -\frac{1+\left(-1\right)^{J_1+J_2+J_3}}{2} {\cal
     G}^{J_1J_2 J_3}_{M_1 M_2 M_3} \frac{
     \CG{J_1}{1}{J_2}{,-1}{J_3}{0}} {
     \CG{J_1}{0}{J_2}{0}{J_3}{0}}, \nonumber \\
     \int d^2\hatn Y_{(J_1M_1)}^{B,a}(\hatn) Y_{(J_2M_2)a}^{E}(\hatn)
     Y_{\left(J_3M_3\right)}(\hatn)
     & = & i\left[\frac{1-(-1)^{J_1+J_2+J_3}}{2}\right] {\cal
     G}^{J_1J_2 J_3}_{M_1 M_2 M_3} \frac{
     \CG{J_1}{1}{J_2}{,-1}{J_3}{0}} {\CG{J_1}{0}{J_2}{0}{J_3}{0}},
\ea
and we also have that
\be  
     Y_{(J_1M_1)a}^{L}(\hatn) Y_{(J_2M_2)}^{E/B,a}(\hatn) = 0.
\ee

From these we obtain that the nonzero TAM-wave overlap integrals
for $J_1+J_2+J_3=$ even are
\begin{eqnarray}
     \int d^{3}x\Psi_{(J_1M_1)a}^{B,k_1}
     (\bfx) \Psi_{(J_2M_2)}^{B,k_2,a}
     (\bfx)\Psi_{(J_3M_3)}^{k_3}(\bfx)  
     & = & -\mathcal{G}_{M_1M_2M_3}^{J_1J_2J_3}
     \frac{\left<J_11J_2,-1|J_30\right>}{\left<J_10J_20|J_30\right>}
     J_{J_1 J_2 J_3}(k_1, k_2, k_3), \label{eqn:3ovl_BBs} \\
     \int d^{3}x \Psi_{(J_1M_1)a}^{E,k_1}
     (\bfx) \Psi_{(J_2M_2)}^{E,k_2,a}
     (\bfx)
     \Psi_{(J_3M_3)}^{k_3}(\bfx) & = & 
     \mathcal{G}_{M_1M_2M_3}^{J_1J_2J_3} \left[
     \frac{\left<J_11J_2,-1|J_30\right>}{\left<J_10J_20|J_30\right>}
     \int      r^2\, dr\,
     j_{J_1,v}^{(E,E)}\left(k_1r\right)j_{J_2,v}^{(E,E)}
     \left(k_2r\right)j_{J_3}\left(k_3r\right)
     \right. \nonumber \\
     & & \left. - \int r^2\,dr\,
     j_{J_1,v}^{(E,L)}\left(k_1r\right)j_{J_2,v}^{(E,L)}
     \left(k_2r\right)j_{J_3}\left(k_3r\right) \right],
\label{eqn:3ovl_EEs}
\end{eqnarray}
and when $J_1+J_2+J_3=\mathrm{odd}$,
\begin{eqnarray}
     \int d^{3}x\Psi_{(J_1M_1)a}^{B,k_1}
     (\bfx) \Psi_{(J_2M_2)}^{E,k_2,a}
     (\bfx)\Psi_{(J_3M_3)}^{k_3} (\bfx) 
     & = &
     \frac{\left<J_11J_2-1|J_30\right>}{\left<J_10J_20|J_30\right>}
     {\cal G}^{J_1 J_2 J_3}_{M_1 M_2 M_3} \int r^2\, dr\,
     j_{J_{1}}\left(k_{1}r\right)j_{J_{2},v}^{(E,E)}
     \left(k_{2}r\right) j_{J_{3},v} \left(k_{3}r\right).\nn\\
\label{eqn:3ovl_BEs}
\end{eqnarray}
Here, we have used the radial functions defined in Eq.~(\ref{eqn:def_jJvec}).
Note that strictly speaking, the Gaunt integral vanishes if
$J_1+J_2+J_3=$ odd, as the Wigner-3j symbol
$\wigner{J_1}{0}{J_2}{0}{J_3}{0}$ vanishes.  In every case where
a Gaunt integral appears with $J_1+J_2+J_3=$ odd, though, it
appears divided by a Clebsch-Gordan coefficient
$\CG{J_1}{0}{J_2}{0}{J_3}{0}$ which also contains the same
Wigner-3j symbol, and thus cancels that in the numerator.  We
choose to write things in this slightly unconventional way to
enable compact, but still unambiguous, expressions.

Eq.~(\ref{eqn:TAMbk_overlaps}), together with the overlap integrals in
Eqs.~(\ref{eqn:3ovl_BBs}), (\ref{eqn:3ovl_EEs}), and
(\ref{eqn:3ovl_BEs}), provide the $B$-$B$-scalar, $E$-$E$-scalar, and
$B$-$E$-scalar TAM-wave bispectra in terms of
$B_{vvs}(k_1,k_2,k_3)$, the Fourier space vector-vector-scalar
bispectrum.

\section{Two transverse-traceless tensors and one scalar}
\label{sec:twoToneS}

We now consider a three-point correlation involving two
transverse-traceless tensors and one scalar.  Such a correlation
arises, for example, in inflation if the tensor modes are
gravitational waves and the scalar field is the curvature
perturbation \cite{Maldacena:2002vr}.

Our goal here will be to calculate the TAM-wave bispectra,
\be
     \VEV{T^{\alpha_1,k_1}_{(J_1M_1)} T^{\alpha_2,k_2}_{(J_2M_2)}
     \phi^{k_3}_{(J_3M_3)}},
\ee
for $\{\alpha_1,\alpha_2\}=\{TE,TB\}$ that correspond to a given
Fourier-space bispectrum.  Here we will take the Fourier-space
tensor-tensor-scalar three-point function to be of the form
\be
     \VEV{ \tilde{T}^{ab}(\bfk_1) \tilde{T}^{cd}(\bfk_2)
     \tilde{\phi}(\bfk_3)} = (2\pi)^3
     \delta_D(\bfk_1+\bfk_2+\bfk_3) g^{ac}g^{bd}
     B_{tts}(k_1,k_2,k_3),
\label{eqn:tts}
\ee
that arises in single-field slow-roll inflation
\cite{Maldacena:2002vr}, in terms of a tensor-tensor-scalar bispectrum
$B_{tts}(k_1,k_2,k_3)$.  The tensor fields $T^{ab}$ are
transverse, $\nabla^a \tilde{T}_{ab}(\bfx)= 0$,
and traceless, $g^{ab}\tilde{T}_{ab}(\bfx)=0$.

We calculate the three-point correlation function of $TE/TB$ TAM 
coefficients directly from Eq.~(\ref{eqn:tensorLEBspherical}): 
\be
     \VEV{T^{\alpha_1,k_1}_{(J_1M_1)} T^{\alpha_2,k_2}_{(J_2M_2)}
     \phi^{k_3}_{(J_3M_3)}} = \int d^2\hatk_1 d^2\hatk_2
     d^2\hatk_3\, Y^{\alpha_1\,\,*}_{(J_1 M_1)ab}(\hatk_1)
     Y^{\alpha_2\,\,*}_{(J_2 M_2)cd}(\hatk_2) Y^{\,\,*}_{(J_3
     M_3)}(\hatk_3) \VEV{\tilde T^{ab}(\bfk_1) \tilde T^{cd}(\bfk_2)
     \tilde \phi(\bfk_3)},
\ee
where $\alpha_1,\alpha_2=\{L,E,B\}$.  Expanding the Dirac delta
function in Eq.~(\ref{eqn:tts}) with Eq.~(\ref{eqn:Diracexpansion}) and  
using the identity,
\be
     \sum_{J'M'}
	  (-i)^{J'} j_{J'}(kr)
	  Y_{(J'M')}^*(\hatn) \int d^2 \hatk\, Y^{\alpha\,\,*}_{(JM)ab}(\hatk)
     Y_{(J'M')}(\hatk) 
	  =(-i)^J \left[\Psi_{(JM) ab}^{\alpha,k}(\bfx)\right]^*,
\ee
we find that
\be
     \VEV{T^{\alpha_1,k_1}_{(J_1M_1)} T^{\alpha_2,k_2}_{(J_2M_2)}
     \phi^{k_3}_{(J_3M_3)}} = 
		(4\pi)^3 (-i)^{J_1+J_2+J_3}
     B_{tts}(k_1,k_2,k_3) 
		\int d^3x\,
     \Psi^{\alpha_1,k_1\,\,*}_{(J_1M_1)}{}^{ab}(\bfx) 
     \Psi^{\alpha_2,k_2\,\,*}_{(J_2 M_2)ab}(\bfx)
     \Psi^{k_3\,\,*}_{(J_3M_3)}(\bfx).
\label{eqn:tensoroverlaps}
\ee
That is, the bispectrum of three TAM-wave coefficients is again
proportional to the overlap of three TAM waves.  

Our task now is to evaluate the overlap integrals,
\begin{equation}
     \int d^{3}x\,
     \Psi_{(J_1M_1)ab}^{\alpha_1,k_1}(\bfx)
     \Psi_{(J_2M_2)}^{\alpha_2,k_2,ab}(\bfx)
     \Psi_{(J_3M_3)}^{k_3}(\bfx),
\end{equation}
for $\alpha_1,\alpha_2=TE,TB$.  With the the decomposition in
Eq.~(94) of Ref.~\cite{Dai:2012}, we only have to calculate
the overlap of three tensor spherical harmonics
\begin{equation}
     \int d^2\hat{n}
     I^{ab,cd}Y_{(J_1M_1)ab}^{\alpha_1}(\hatn)
     Y_{(J_2M_2)cd}^{\alpha_2}(\hatn)
     Y_{\left(J_3M_3\right)}(\hatn),
\end{equation}
for $\alpha_1,\alpha_2=L,VE/VB,TE/TB$.
Here the constant tensor $I^{ab,cd}$ is the identity tensor acting on symmetric
traceless tensors. In Appendix~\ref{app:3Ylms_LEB_tensor}, we
calculate these overlaps in the helicity basis.  We then use the
relations,
\begin{equation}
     Y_{(JM)ab}^{\pm2}=\frac{1}{\sqrt{2}}
     \left(Y_{(JM)ab}^{TE}\pm i Y_{(JM)ab}^{TB}\right),\quad
     Y_{(JM)ab}^{\pm1}=\frac{1}{\sqrt{2}}
     \left(Y_{(JM)ab}^{VE}\pm iY_{(JM)ab}^{VB}\right), \quad
     Y_{(JM)ab}^{0}=Y_{(JM)ab}^{L},
\end{equation}
to find that 
\ba
     \int d^2\hatn\,Y_{(J_1M_1)}^{L,ab}(\hatn)
     Y_{(J_2M_2)ab}^{L}(\hatn) Y_{\left(J_3M_3\right)}(\hatn)  
     \label{eqn:3ovl_YLYLYs}
     &=  & \mathcal{G}^{J_1J_2J_3}_{M_1M_2M_3}, \\
     \int d^2\hatn\, Y_{(J_1M_1)}^{VB,ab}(\hatn)
     Y_{(J_2M_2)ab}^{VB}(\hatn) Y_{\left(J_3M_3\right)}(\hatn) 
     \label{eqn:3ovl_YVBYVBYs}
     &=& \int d^2\hatn Y_{(J_1M_1)}^{VE,ab}(\hatn)
     Y_{(J_2M_2)ab}^{VE}(\hatn)
     Y_{\left(J_3M_3\right)}(\hatn) \nonumber \\
     & = & -\frac{1+\left(-1\right)^{J_1+J_2+J_3}}{2}
     \mathcal{G}^{J_1J_2J_3}_{M_1M_2M_3} \frac{
     \CG{J_1}{1}{J_2}{,-1}{J_3}{0} }{\CG{J_1}{0}{J_2}{0}{J_3}{0}
     }, \\
     \int d^2\hatn\, Y_{(J_1M_1)}^{VB,ab}
     Y_{(J_2M_2)ab}^{VE}(\hatn)
     Y_{\left(J_3M_3\right)}(\hatn)    
     \label{eqn:3ovl_YVBYVEYs}
     &=&  i\frac{1-\left(-1\right)^{J_1+J_2+J_3}}{2} 
     \mathcal{G}^{J_1J_2J_3}_{M_1M_2M_3} \frac{
     \CG{J_1}{1}{J_2}{,-1}{J_3}{0} }{\CG{J_1}{0}{J_2}{0}{J_3}{0}
     }, \\
     \int d^2\hat{n}Y_{(J_1M_1)}^{TB,ab}(\hatn)
     Y_{(J_2M_2)ab}^{TB}(\hatn) Y_{\left(J_3M_3\right)}(\hatn) 
     \label{eqn:3ovl_YTBYTBYs}
     &=& \int d^2\hatn Y_{(J_1M_1)}^{TE,ab}(\hatn)
     Y_{(J_2M_2)ab}^{TE}(\hatn)
     Y_{\left(J_3M_3\right)}(\hatn) \nonumber \\
     & = & \frac{1+\left(-1\right)^{J_1+J_2+J_3}}{2}
     \mathcal{G}^{J_1J_2J_3}_{M_1M_2M_3} \frac{
     \CG{J_1}{2}{J_2}{,-2}{J_3}{0} }{\CG{J_1}{0}{J_2}{0}{J_3}{0}
     }, \\
     \int d^2\hatn\, Y_{(J_1M_1)}^{TB,ab}(\hatn)
     Y_{(J_2M_2)ab}^{TE}(\hatn) Y_{\left(J_3M_3\right)}(\hatn) 
     \label{eqn:3ovl_YTBYTEYs}
     &=&  -i\frac{1-\left(-1\right)^{J_1+J_2+J_3}}{2} 
     \mathcal{G}^{J_1J_2J_3}_{M_1M_2M_3} \frac{
     \CG{J_1}{2}{J_2}{,-2}{J_3}{0} }{\CG{J_1}{0}{J_2}{0}{J_3}{0}
     }.
\ea

We now use Eq.~(94) of Ref.~\cite{Dai:2012}, along with
the orthonormality conditions,
\begin{equation}
     Y_{(J_1M_1)ab}^{L}(\hatn)Y_{(J_2M_2)}^{VE/VB,ab}(\hatn)
     =0,\quad Y_{(J_1M_1)ab}^{L}(\hatn)
     Y_{(J_2M_2)}^{TE/TB,ab}(\hatn) =0,\quad
     Y_{(J_1M_1)ab}^{VE/VB}(\hatn)
     Y_{(J_2M_2)}^{TE/TB,ab}(\hatn) =0,
\end{equation}
to find that for $J_1+J_2+J_3=\mathrm{even}$,
\begin{eqnarray}
&&     \int d^{3}x\,
     \Psi_{(J_1M_1)ab}^{TB,k_1}(\bfx)
     \Psi_{(J_2M_2)}^{TB,k_2,ab}(\bfx)
     \Psi_{(J_3M_3)}^{k_3}(\bfx) =
     \mathcal{G}_{M_1M_2M_3}^{J_1J_2J_3} \int r^2 dr
     j_{J_3}(k_3r) \left[
     \frac{\left<J_11J_2,-1|J_30\right>}{\left<J_10J_20|J_30\right>}
     j_{J_1,t}^{(TB,VB)}(k_1r)j_{J_2,t}^{(TB,VB)}(k_2r) \right. \nonumber \\
     &&\qquad\qquad\qquad\qquad \left. -
     \frac{\left<J_12J_2,-2|J_30\right>}{\left<J_10J_20|J_30\right>}
     j_{J_1,t}^{(TB,TB)}(k_1r)j_{J_2,t}^{(TB,TB)}(k_2r) \right],
     \label{eqn:3ovl_TBTBs} \\
     &&\int d^{3}x\, \Psi_{(J_1M_1)ab}^{TE,k_1}(\bfx)
     \Psi_{(J_2M_2)}^{TE,k_2,ab}(\bfx)
     \Psi_{(J_3M_3)}^{k_3}(\bfx) = \mathcal{G}_{M_1M_2M_3}^{J_1J_2J_3}
     \int r^2 dr  j_{J_3}(k_3r) \Biggl[
     j_{J_1,t}^{(TE,L)}(k_1r)j_{J_2,t}^{(TE,L)}(k_2r) \nonumber \\
     &&\qquad\qquad      -
     \frac{\left<J_11J_2,-1|J_30\right>}{\left<J_10J_20|J_30\right>}
     j_{J_1,t}^{(TE,VE)}(k_1r)j_{J_2,t}^{(TE,VE)}(k_2r)
      + \frac{\left<J_12J_2,-2|J_30\right>}
     {\left<J_10J_20|J_30\right>}
     j_{J_1,t}^{(TE,TE)}(k_1r)j_{J_2,t}^{(TE,TE)}(k_2r)\Biggr],
\label{eqn:3ovl_TETEs}
\end{eqnarray}
and for $J_1+J_2+J_3=\mathrm{odd}$, 
\begin{eqnarray}
     \int d^{3}x\, \Psi_{(J_1M_1)ab}^{TB,k_1}(\bfx)
     \Psi_{(J_2M_2)}^{TE,k_2,ab}(\bfx) \Psi_{(J_3M_3)}^{k_3}(\bfx)
     &=& \mathcal{G}_{M_1M_2M_3}^{J_1J_2J_3}
     \int r^2drj_{J_3}\left(k_3r\right) \Biggl[- \frac{
     \CG{J_1}{1}{J_2}{,-1}{J_3}{0} 
     }{\CG{J_1}{0}{J_2}{0}{J_3}{0}}
     j_{J_1,t}^{(TB,VB)}\left(k_1r\right)j_{J_2,t}^{(TE,VE)}\left(k_2r\right) 
     \nonumber \\ 
     & & + \frac{ \CG{J_1}{2}{J_2}{,-2}{J_3}{0}
     }{\CG{J_1}{0}{J_2}{0}{J_3}{0}}
     j_{J_1,t}^{(TB,TB)}\left(k_1r\right)
     j_{J_2,t}^{(TE,TE)}\left(k_2r\right)\Biggr].
\label{eqn:3ovl_TBTEs}
\end{eqnarray}
Here, we have used the radial functions defined in Eq.~(\ref{eqn:def_jJten1}) and Eq.~(\ref{eqn:def_jJten2}).
Eq.~(\ref{eqn:tensoroverlaps}), together with the overlap
integrals in Eqs.~(\ref{eqn:3ovl_TBTBs}),
(\ref{eqn:3ovl_TETEs}), and (\ref{eqn:3ovl_TBTEs}), provides the
$TB$-$TB$-scalar, $TE$-$TE$-scalar, and $TB$-$TE$-scalar
TAM-wave bispectra in terms of $B_{tts}(k_1,k_2,k_3)$, the
Fourier-space tensor-tensor-scalar bispectrum.

\section{One symmetric traceless tensor and two scalars (or
two longitudinal vectors)}
\label{sec:oneTtwoS}

We now calculate the three-point correlation that involves one
symmetric traceless tensor and two scalars. The traceless tensor
may be a longitudinal ($L$) mode, one of the two
vector modes ($VE$ or $VB$), or one of the two transverse-tensor
($TE$ or $TB$) modes. The three-dimensional galaxy-survey
observables of such a three-point tensor-scalar-scalar
correlation were studied in
Ref.~\cite{Jeong:2012df}.  Moreover, a tensor-scalar-scalar
correlation between inflationary gravitational waves (the tensor
mode) and the primordial curvature perturbation (the scalar
mode) arises generically during inflation
\cite{Maldacena:2002vr}.  The TAM-wave bispectra we present
here may be useful in calculating full-sky observables
associated with these couplings.

To be precise, we consider here a tensor-scalar-scalar
bispectrum $B_{sst}(k_1,k_2,k_3)$ defined by
\be
\VEV{
\widetilde{\nabla^a \phi}(\bfk_1)
\widetilde{\nabla^b \phi}(\bfk_2)
\tilde{T}^{cd}(\bfk_3)
}
=
(2\pi)^3
\delta^D(\bfk_1+\bfk_2+\bfk_3)
g^{ac}g^{bd}
k_1 k_2 B_{sst}(k_1,k_2,k_3).
\label{eqn:3DbispSST}
\ee
Given that $V^a= \nabla^a \phi$ is a longitudinal-vector field, we
can also write this three-point function as a
vector-vector-tensor bispectrum $B_{vvt}(k_1,k_2,k_3)$, where
here ``vector'' is a longitudinal vector, of the form,
\be
\VEV{
\tilde{V}^{a}(\hatk_1)
\tilde{V}^{b}(\hatk_2)
\tilde{T}^{cd}(\hatk_3)
}
=
(2\pi)^3
\delta^D(\bfk_1+\bfk_2+\bfk_3)
g^{ac}g^{bd}
B_{vvt}(k_1,k_2,k_3).
\label{eqn:3DbispVVT}
\ee

The three-point function of one TAM coefficient
and two scalar TAM coefficients can be calculated from
\be
\VEV{V^{L,k_1}_{(J_1M_1)} V^{L,k_2}_{(J_2M_2)} T^{\alpha,k_3}_{(J_3M_3)}}
=
(4\pi)^3(-i)^{J_1+J_2+J_3}
B_{vvt}(k_1,k_2,k_3)
\int d^3x
\Psi_{(J_1M_1)}^{L,k_1\,\, a}(\bfx)
\Psi_{(J_2M_2)}^{L,k_2\,\,b}(\bfx)
\Psi_{(J_3M_3)ab}^{\alpha,k_3}(\bfx),
\label{eqn:vvtoverlap}
\ee
and using Eq.~(\ref{eqn:3DbispSST}) and $\phi_{JM}^k = iV_{(JM)}^{L,k}/k$,
\be
\VEV{\phi^{k_1}_{(J_1M_1)} \phi^{k_2}_{(J_2M_2)} T^{\alpha,k_3}_{(J_3M_3)}}
=
- (4\pi)^3(-i)^{J_1+J_2+J_3}
B_{sst}(k_1,k_2,k_3)
\int d^3x
\Psi_{(J_1M_1)}^{L,k_1\,\, a}(\bfx)
\Psi_{(J_2M_2)}^{L,k_2\,\, b}(\bfx)
\Psi_{(J_3M_3)ab}^{\alpha,k_3}(\bfx).
\label{eqn:sstoverlap}
\ee
That is, we only need to calculate the overlap,
\be
\label{eq:overlap-sst}
\int d^{3}x \Psi_{(J_1M_1)}^{L,k_1\,\,a}(\bfx)
\Psi_{(J_2M_2)}^{L,k_2\,\,b}(\bfx)\Psi_{(J_3M_3)ab}^{\alpha,k_3}(\bfx)
=
\int d^{3}x\left[\frac{i}{k_1}\nabla^{a}\Psi_{(J_1M_1)}^{k_1}(\bfx)\right]\left[\frac{i}{k_2}\nabla^{b}\Psi_{(J_2M_2)}^{k_2}(\bfx)\right]\Psi_{(J_3M_3)ab}^{\alpha,k_3}(\bfx),
\ee
for $\alpha=L,VE,VB,TE,TB$. Below we discuss the three cases
where the traceless tensor is longitudinal ($L$), vectorial
($VE/VB$), or transverse ($TE/TB$).

\subsection{The transverse-traceless components}

We start with transverse-traceless tensors, $\alpha=TE,TB$. 
Since the transverse-traceless fields are divergence-free, we can
integrate by parts in the integral in
Eq.~(\ref{eq:overlap-sst}) to get,
\begin{equation}
\frac{1}{k_1k_2}\int d^{3}x\Psi_{(J_1M_1)}^{k_1}(\bfx)\left[\nabla_{a}\nabla_{b}\Psi_{(J_2M_2)}^{k_2}(\bfx)\right]\Psi_{(J_3M_3)}^{\alpha,k_1,ab}(\bfx).
\end{equation}
We next use the definition~\cite{Dai:2012},
\begin{equation}
\label{eq:longitudinal-TAM-operator}
\Psi_{(JM)ab}^{L,k}(\bfx)=\sqrt{\frac{3}{2}}\left(\frac{1}{k^2}\nabla_{a}\nabla_{b}+\frac{1}{3}g_{ab}\right)\Psi_{(JM)}^{k}(\bfx),
\end{equation}
of longitudinal-tensor TAM waves,
and the fact that the tensor mode is traceless, 
to rewrite the three-wavefunction overlap as,
\ba
\int d^{3}x \Psi_{(J_1M_1)}^{L,k_1\,\,a}(\bfx)
\Psi_{(J_2M_2)}^{L,k_2\,\,b}(\bfx)\Psi_{(J_3M_3)ab}^{\alpha,k_3}(\bfx)
=
\sqrt{\frac{1}{6}}
\left[
\frac{k_2}{k_1}\int d^{3}x\Psi_{(J_1M_1)}^{k_1}(\bfx)\Psi_{(J_2M_2)ab}^{L,k_2}(\bfx)\Psi_{(J_3M_3)}^{\alpha,k_3,\,ab}(\bfx)
+(1\leftrightarrow2)
\right],
\ea
for $\alpha=TE,TB$.
What we now need to calculate is the overlap of a transverse tensor, a
longitudinal tensor, and a scalar. 
Using 
Eq.~(\ref{eqn:tensorprojections})
and
Eqs.~(\ref{eqn:3ovl_YLYLYs})--(\ref{eqn:3ovl_YTBYTEYs}),
we find that when
$J_1+J_2+J_3=\mathrm{even}$,
\begin{eqnarray}
    && \int d^{3}x\Psi_{(J_1M_1)}^{L,k_1\,\,a}(\bfx)
     \Psi_{(J_2M_2)}^{L,k_2\,\,b}(\bfx)\Psi_{(J_3M_3)ab}^{TE,k_3}(\bfx)
     = \frac{1}{\sqrt{6}} \mathcal{G}^{J_3J_2J_1}_{M_3M_2M_1}
     \biggl\{
     \frac{k_2}{k_1}\int r^2drj_{J_1}\left(k_1r\right)
     \left[
     j_{J_2,t}^{(L,L)}\left(k_2r\right)j_{J_3,t}^{(TE,L)}\left(k_3r\right)
     \right. \vs
     &&\qquad\left.-\frac{ \CG{J_3}{1}{J_2}{,-1}{J_1}{0}}{
     \CG{J_3}{0}{J_2}{0}{J_1}{0}}
     j_{J_2,t}^{(L,VE)}\left(k_2r\right)j_{J_3,t}^{(TE,VE)}\left(k_3r\right)
     + \frac{ \CG{J_3}{2}{J_2}{,-2}{J_1}{0}}{ \CG{J_3}{0}{J_2}{0}{J_1}{0}}
     j_{J_2,t}^{(L,TE)}\left(k_2r\right)
     j_{J_3,t}^{(TE,TE)}\left(k_3r\right)\right] +
     (1\leftrightarrow2)
     \biggl\},\nn\\
\label{eqn:ssteven}
\end{eqnarray}
and when $J_1+J_2+J_3=\mathrm{odd}$,
\begin{eqnarray}
    && \int d^{3}x\Psi_{(J_1M_1)}^{L,k_1\,\,a}(\bfx)
     \Psi_{(J_2M_2)}^{L,k_2\,\,b}(\bfx)\Psi_{(J_3M_3)ab}^{TB,k_3}(\bfx)
     =\frac{1}{\sqrt{6}} \mathcal{G}^{J_3J_2J_1}_{M_3M_2M_1}
     \biggl\{
     \frac{k_2}{k_1}\int r^2drj_{J_1}\left(k_1r\right)
     \vs
      & & \qquad\times\left[- \frac{ \CG{J_3}{1}{J_2}{,-1}{J_1}{0}}{
     \CG{J_3}{0}{J_2}{0}{J_1}{0}}
      j_{J_2,t}^{(L,VE)}\left(k_2r\right)j_{J_3,t}^{(TB,VB)}\left(k_3r\right)
      +
      \frac{ \CG{J_3}{2}{J_2}{,-2}{J_1}{0}}{ \CG{J_3}{0}{J_2}{0}{J_1}{0}}
      j_{J_2,t}^{(L,TE)}\left(k_2r\right)
      j_{J_3,t}^{(TB,TB)}\left(k_3r\right)\right]
      +    (1\leftrightarrow2) \biggl\}.\nn\\
\label{eqn:sstodd}
\end{eqnarray}
Here, again, we use the radial functions defined 
in Eq.~(\ref{eqn:def_jJten1}) and Eq.~(\ref{eqn:def_jJten2}).

\subsection{The vector components}

Next we consider the two vector modes, $\alpha=VE,VB$. The
TAM waves for the vector components of the symmetric trace-free
tensor can be related to the TAM waves for transverse-vector
fields~\cite{Dai:2012} through
\be
\frac{i}{k}
\nabla_a \Psi_{(JM)}^{VB,ab}(\bfx)
=
-\frac{1}{\sqrt{2}} \Psi_{(JM)}^{B,b}(\bfx)
,\quad
\frac{i}{k}
\nabla_a \Psi_{(JM)}^{VE,ab}(\bfx)
=
-\frac{1}{\sqrt{2}} \Psi_{(JM)}^{E,b}(\bfx).
\ee
Since the $\Psi_{(JM)ab}^{TE/TB,k}$ are traceless, we can
do the overlap integral in Eq.~(\ref{eq:overlap-sst}) by
integrating by parts to obtain
\ba
\int d^{3}x\Psi_{(J_1M_1)a}^{L,k_1}(\bfx)\Psi_{(J_2M_2)b}^{L,k_2}(\bfx)\Psi_{(J_3M_3)}^{\alpha,k_3,ab}(\bfx)
&=&
\sqrt{\frac{1}{6}}
\left[
\frac{k_2}{k_1}\int d^{3}x\Psi_{(J_1M_1)}^{k_1}(\bfx)\Psi_{(J_2M_2)ab}^{L,k_2}(\bfx)\Psi_{(J_3M_3)}^{\alpha,k_3,\,ab}(\bfx)
+(1\leftrightarrow2)
\right]
\nonumber\\
&&+
\frac{1}{\sqrt{8}}
\left[
\frac{k_3}{k_1}
\int d^3 x  
\Psi_{(J_1M_1)}^{k_1}(\bfx)
\Psi_{(J_2M_2)b}^{L,k_2}(\bfx)
\Psi_{(J_3M_3)}^{\alpha,k_3,b}(\bfx)
+(1\leftrightarrow2)
\right],
\ea
where $\alpha=TE,TB$ for tensor-valued TAM waves, and
$\alpha=E,B$ for vector-valued TAM waves.
Following the same techniques as in previous sections, we find 
for $J_1+J_2+J_3=$even, 
\ba
\label{eqn:sstveven}
&&\int d^{3}x\Psi_{(J_1M_1)a}^{L,k_1}(\bfx)\Psi_{(J_2M_2)b}^{L,k_2}(\bfx)\Psi_{(J_3M_3)}^{VE,k_3,ab}(\bfx)
=
\sqrt{\frac{1}{6}}
\mathcal{G}_{M_3M_2M_1}^{J_3J_2J_1}
\Biggl[
\frac{k_2}{k_1}
\int r^2 dr j_{J_1}(k_1r)
\biggl\{
j_{J_2,t}^{(L,L)}(k_2r)j_{J_3,t}^{(VE,L)}(k_3r)
\nonumber\\
&&\quad\quad-
\frac{\langle J_31J_2,-1|J_10\rangle}{\langle J_30J_20|J_10\rangle}
j_{J_2,t}^{(L,VE)}(k_2r)j_{J_3,t}^{(VE,VE)}(k_3r)
+
\frac{\langle J_32J_2,-2|J_10\rangle}{\langle J_30J_20|J_10\rangle}
j_{J_2,t}^{(L,TE)}j_{J_3,t}^{(VE,TE)}
\biggl\}
+(1\leftrightarrow2)
\Biggl]
\nonumber\\
&&-
\frac{1}{\sqrt{8}}
\mathcal{G}_{M_3M_2M_1}^{J_3J_2J_1}
\Biggl[
\frac{k_3}{k_1}
\int r^2 dr j_{J_1}(k_1r)
\biggl\{
j_{J_2,v}^{(L,L)}(k_2r)
j_{J_3,v}^{(E,L)}(k_3r)
-
\frac{\langle J_31J_2,-1|J_10\rangle}{\langle J_30J_20|J_10\rangle}
j_{J_2,v}^{(L,E)}(k_2r)
j_{J_3,v}^{(E,E)}(k_3r)
\biggl\}
+(1\leftrightarrow2)
\Biggl],\nn\\
\ea
and for $J_1+J_2+J_3=$odd, 
\ba
\label{eqn:sstvodd}
&&\int d^{3}x\Psi_{(J_1M_1)a}^{L,k_1}(\bfx)\Psi_{(J_2M_2)b}^{L,k_2}(\bfx)\Psi_{(J_3M_3)}^{VB,k_3,ab}(\bfx)
=
\sqrt{\frac{1}{6}}
\mathcal{G}_{M_3M_2M_1}^{J_3J_2J_1}
\Biggl[
\frac{k_2}{k_1}
\int r^2 dr j_{J_1}(k_1r)
\nonumber\\
&&\quad\quad\times
\biggl\{
-
\frac{\langle J_31J_2,-1|J_10\rangle}{\langle J_30J_20|J_10\rangle}
j_{J_2,t}^{(L,VE)}(k_2r)j_{J_3,t}^{(VB,VB)}(k_3r)
+
\frac{\langle J_32J_2,-2|J_10\rangle}{\langle J_30J_20|J_10\rangle}
j_{J_2,t}^{(L,TE)}j_{J_3,t}^{(VB,TB)}
\biggl\}
+(1\leftrightarrow2)
\Biggl]
\nonumber\\
&&+
\frac{1}{\sqrt{8}}
\mathcal{G}_{M_3M_2M_1}^{J_3J_2J_1}
\Biggl[
\frac{k_3}{k_1}
\frac{\langle J_31J_2,-1|J_10\rangle}{\langle J_30J_20|J_10\rangle}
\int r^2 dr j_{J_1}(k_1r)
j_{J_2,v}^{(L,E)}(k_2r)
j_{J_3}(k_3r)
+(1\leftrightarrow2)
\Biggl].
\ea
Again, we have used the radial functions defined 
in Eq.~(\ref{eqn:def_jJten1}) and Eq.~(\ref{eqn:def_jJten2}).

\subsection{Longitudinal part}

Finally, we consider the longitudinal mode of the traceless
tensor field. Once again, we can write the longitudinal
tensor-valued TAM wave function
$\Psi^{L,k}_{(JM)ab}(\mathbf{x})$ in terms of differential
operators acting on scalar-valued TAM waves
$\Psi^{k}_{(JM)}(\mathbf{x})$, as shown in
Eq.~(\ref{eq:longitudinal-TAM-operator}). We can also put the
two longitudinal vector-valued TAM wave functions into this
operator form. Integrating by parts 
allows us to move around those gradient operators, and we
re-write the overlap of three TAM waves as
\begin{align}
\int d^{3}x\Psi_{(J_1M_1)}^{L,k_1,a}(\bfx)\Psi_{(J_2M_2)}^{L,k_2,b}(\bfx)\Psi_{(J_3M_3)ab}^{L,k_3}(\bfx)=&\frac{1}{\sqrt{6}}\int d^{3}x\left\{ \frac{k_2}{k_1}\Psi_{(J_1M_1)}^{k_1}(\bfx)\Psi_{(J_2M_2)}^{L,k_2,ab}(\bfx)\Psi_{(J_3M_3)ab}^{L,k_3}(\bfx)\right.\nn\\
&\left.+\frac{k_3}{k_1}\Psi_{(J_1M_1)}^{k_1}(\bfx)\Psi_{(J_2M_2)}^{L,k_2,a}(\bfx)\Psi_{(J_3M_3)a}^{L,k_3}(\bfx)+\left(1\leftrightarrow2\right)\right\}.
\end{align}
The first term is the overlap of two longitudinal tensors and
one scalar, and the second term is the overlap of two longitudinal
vectors and one scalar. The integrals are non-zero only if
$J_1+J_2+J_3=$ even, and the end result is  
\begin{align}
\label{eqn:sstleven}
\int d^{3}&x\Psi_{(J_1M_1)}^{L,k_1,a}(\bfx)\Psi_{(J_2M_2)}^{L,k_2,b}(\bfx)\Psi_{(J_3M_3)ab}^{L,k_3}(\bfx)=\frac{1}{\sqrt{6}}\mathcal{G}_{M_3M_2M_1}^{J_3J_2J_1}\left\{ \int r^2drj_{J_1}\left(k_1r\right)\left[\frac{k_2}{k_1}\left(j_{J_2,t}^{(L,L)}\left(k_2r\right)j_{J_3,t}^{(L,L)}\left(k_3r\right)\right.\right.\right.\nn\\
&\left.\left.\left.-\frac{\langle J_31J_2,-1|J_10\rangle}{\langle J_30J_20|J_10\rangle}j_{J_2,t}^{(L,VE)}\left(k_2r\right)j_{J_3,t}^{(L,VE)}\left(k_3r\right)+\frac{\langle J_32J_2,-2|J_10\rangle}{\langle J_30J_20|J_10\rangle}j_{J_2,t}^{(L,TE)}\left(k_2r\right)j_{J_3,t}^{(L,TE)}\left(k_3r\right)\right) \right.\right.\nn\\
&\left.\left. -\frac{k_3}{k_1}\left(j_{J_2,v}^{(L,L)}\left(k_2r\right)j_{J_3,v}^{(L,L)}\left(k_3r\right)-\frac{\langle J_31J_2,-1|J_10\rangle}{\langle J_30J_20|J_10\rangle}j_{J_2,v}^{(L,E)}\left(k_2r\right)j_{J_3,v}^{(L,E)}\left(k_3r\right)\right) \right]  +\left(1\leftrightarrow2\right)\right\},
\end{align}
where the various radial functions have been defined 
in Eq.~(\ref{eqn:def_jJten1}) in Eq.~(\ref{eqn:def_jJten2}).

We have thus calculated the overlaps of a traceless-tensor TAM
wave and two scalar (or longitudinal-vector) TAM waves for the
longitudinal, vector, and transverse-tensor components. These
results, when inserted into Eq.~(\ref{eqn:vvtoverlap}) or
Eq.~(\ref{eqn:vvtoverlap}), give the TAM bispectra for a
traceless tensor field with either two longitudinal fields or
two scalar fields.

\section{Conclusions}
\label{sec:conclusion}

In this paper, we have calculated several bispectra for scalar,
vector, and tensor fields in the total-angular-momentum
formalism.  We began with the scalar-scalar-scalar bispectrum.
We then considered an example of a vector-vector-scalar (where
here vector is a transverse vector) bispectrum and a
tensor-tensor-scalar (where here tensor is a
transverse-traceless tensor).  The vector-vector-scalar
bispectrum is of the form that may arise from the correlation of
a magnetic field with a scalar field \cite{Caldwell:2011ra},
while the tensor-tensor-scalar correlation is precisely the same
form that arises in inflation \cite{Maldacena:2002vr}.  We
obtained bispectra for TAM waves in the $E/B$ basis (for vector
fields) and the $TE/TB$ basis (for the tensor fields) through
intermediate steps that involved the TAM-wave helicity basis.
We then moved on to calculate the TAM-wave three-point function
for a tensor-scalar-scalar correlation that comes from a
tensor-scalar-scalar bispectrum of precisely the same form that
arises in inflation \cite{Maldacena:2002vr}.  For completeness
(and to follow through in Ref.~\cite{inprogress} on CMB
signatures of correlations of the form considered in
Ref.~\cite{Jeong:2012df}), we have also considered the bispectra
for two scalars and either the longitudinal of vector components
of the traceless tensor field.

\begin{table}[htbp]
\begin{center}
\begin{tabular}{|c|c|c|} 
\hline
type of correlation &  Fourier-space bispectrum & TAM-wave
result \\ \hline
scalar-scalar-scalar &  (\ref{eqn:Fourierbispectrum}) &
(\ref{eqn:scalarfinal}) \\ \hline
vector-vector-scalar (transverse vectors) &  (\ref{eqn:vvs}) &
(\ref{eqn:TAMbk_overlaps}) and (\ref{eqn:3ovl_BBs})--(\ref{eqn:3ovl_BEs})
\\ \hline
tensor(T)-tensor(T)-scalar & (\ref{eqn:tts}) &
(\ref{eqn:tensoroverlaps}) and
(\ref{eqn:3ovl_TBTBs})--(\ref{eqn:3ovl_TBTEs}) \\ \hline
tensor(T)-scalar-scalar & (\ref{eqn:3DbispSST}) &
(\ref{eqn:sstoverlap}), (\ref{eqn:ssteven}), and
(\ref{eqn:sstodd}) \\ \hline
tensor(T)-vector-vector (longitudinal vectors) & (\ref{eqn:3DbispVVT}) &
(\ref{eqn:vvtoverlap}), (\ref{eqn:ssteven}), and
(\ref{eqn:sstodd}) \\ \hline
tensor(V)-scalar-scalar & (\ref{eqn:3DbispSST}) &
(\ref{eqn:sstoverlap}), (\ref{eqn:sstveven}), and
(\ref{eqn:sstvodd}) \\ \hline
tensor(V)-vector-vector (longitudinal vectors) & (\ref{eqn:3DbispVVT}) &
(\ref{eqn:vvtoverlap}), (\ref{eqn:sstveven}), and
(\ref{eqn:sstvodd}) \\ \hline
tensor(L)-scalar-scalar & (\ref{eqn:3DbispSST}) &
(\ref{eqn:sstoverlap}) and (\ref{eqn:sstleven}) \\ \hline
tensor(L)-vector-vector (longitudinal vectors) & (\ref{eqn:3DbispVVT}) &
(\ref{eqn:vvtoverlap}) and (\ref{eqn:sstleven}) \\ \hline
\end{tabular}
\caption{List of the types of three-point functions we consider, the
     equation where the Fourier-space bispectrum we consider is
     defined, and the equations that contain the central results
     for the TAM-wave three-point functions.  We distinguish
     between longitudinal and transverse vectors. The labels $T$, $V$ and $L$ refer to 
     transverse tensorial, transverse vectorial and longitudinal part of
     a traceless tensor field respectively.}
\end{center}
\end{table}

In the plane-wave formalism, the three-point
correlation in Fourier space is parametrized by a 
bispectrum $B(k_1,k_2,k_3)$ that depends on the three
wavenumber magnitudes $k_1,k_2,k_3$ only, multiplied by a
``momentum-conserving'' Dirac delta function, a consequence of
statistical homogeneity.  In the TAM formalism, generically the three-point
correlation is parametrized by exactly the same bispectrum, but
multiplied by an ``angular-momentum-conserving'' Clebsch-Gordan
coefficient (which we write in shorthand as a Gaunt factor), a
consequence of statistical isotropy (the Wigner-Eckart theorem),
and an integral over radial profiles which does not depend on the azimuthal quantum numbers.  Statistical homogeneity
along the radial direction is encoded in the integral over the
radial profiles. 

Since the $E/B$ and $TE/TB$ TAM
waves have definite parity, parity conservation is manifest
in the bispectrum (determined by whether $J_1+J_2+J_3$ is even
or odd).  This is to be contrasted with the plane-wave
formalism, where the two linear polarizations of a transverse-vector or a transverse-traceless-tensor are identified with each other by a simple
rotation about the direction of wave propagation.  In the TAM
formalism, the $E/B$ modes have different geometrical factors and
different radial integrals in the bispectrum, despite the fact
that their power spectra must coincide by statistical isotropy
\cite{Dai:2012}. This means that $E/B$ modes correlate to other
fields differently when cubic interactions are taken into
consideration.  They thus have different bispectra.

Our survey of bispectra was not at all exhaustive.  We showed in
Section~\ref{sec:fourierspace}, for example, that
vector-vector-scalar correlations can be parametrized, assuming
parity invariance, in terms of two bispectra, and we then
considered one of those.  There may likewise be additional types
of bispectra involving tensor fields and beyond those that we
considered here.  There are also additional forms for bispectra
that may arise if we allow for parity violation, and in this
case, the helicity-basis TAM waves we developed in
Ref.~\cite{Dai:2012} should be particularly appropriate.

The value of the TAM-wave bispectra we have discussed here
becomes apparent when we realize that many cosmological
measurements are performed on the full sky, or over wide-angle
surveys.  Angular correlations are then decomposed into angular
power spectra $C_J$ (usually written as $C_l$) parametrized by
multipoles $J$.  The advantage of TAM waves is that once the
proper scalar, vector, or tensor TAM waves are identified,
rotational symmetry guarantees that the observable $C_J$ will
receive contributions only from TAM waves with that same $J$.
In this sense, TAM waves provide a more natural choice of basis
functions than the conventionally used plane waves.  The
advantage for three-point functions is that once the proper TAM
waves are identified, the results we have derived in this paper
provide the TAM-wave angular bispectra in terms of the more
commonly seen Fourier-space bispectra.  Since observable angular
bispectra will be obtained from projections of the TAM-wave
bispectra, the $J_1,M_1,J_2,M_2,J_3,M_3$ dependences of the angular
bispectra can be read off directly from the results we have
presented here. For calculating CMB observables from bispectra involving vector or tensor fields,
the TAM approach is particularly powerful in that equivalent results can be obtained without any Wigner-6$j$ or Wigner-9$j$ symbol, 
nor does any additional summation over the orbital angular momentum arise, 
as opposed to using plane waves.

One example of the utility of this formalism will be provided in
forthcoming work \cite{inprogress} where we calculate the
bipolar power spectrum of the cosmic microwave background that
arises from scalar, vector, and tensor distortions to the local
density two-point autocorrelation function of the form discussed
in Ref.~\cite{Jeong:2012df}.  In other work \cite{Trivedi}
we will show how these TAM-wave bispectra can be used to
construct estimators from wide-angle galaxy surveys (including
redshift-space distortions) for the types of distortions
considered in Ref.~\cite{Jeong:2012df}.


\begin{acknowledgments}
LD acknowledges the support of the Rowland Research Fund.  This
work was supported by DoE SC-0008108 and NASA NNX12AE86G.
\end{acknowledgments}

\appendix

\section{TAM-wave overlap integrals in the helicity basis}
\label{app:3Ylms_LEB}
In this Section, we calculate the overlap of three TAM waves in
the helicity basis. We first calculate the overlap of two
transverse-vector TAM waves and one scalar TAM wave in  
Section~\ref{app:3Ylms_LEB_vector} and then the overlap of two
transverse-tensor TAM waves and one scalar TAM wave in
Section~\ref{app:3Ylms_LEB_tensor}.

Our starting point will be the overlap \cite{Hu:2000ee},
\be
     \int d^2 \nhat \,
     {}_{s_1}Y_{(l_1m_1)}(\nhat)
     {}_{s_2}Y_{(l_2m_2)}(\nhat)
     {}_{s_3}Y_{(l_3m_3)}(\nhat)
     =
     (-1)^{l_1+l_2+l_3+s_3} \mathcal{G}^{l_1l_2l_3}_{m_1m_2m_3}
     \frac{\CG{l_1}{s_1}{l_2}{s_2}{l_3}{-s_3}}{\CG{l_1}{0}{l_2}{0}{l_3}{0}},
\label{eqn:three_sYlm}
\ee
of three spin-$s$ spherical harmonics, which holds for
$s_1+s_2+s_3=0$.

\subsection{Two transverse vector TAM-waves and one scalar TAM wave}
\label{app:3Ylms_LEB_vector}
Let us first calculate 
\begin{equation}
     \int
     d^2\hatn g^{ab}Y_{(J_1M_1)a}^{\lambda_1}
     \left(\hatn\right)
     Y_{(J_2M_2)b}^{\lambda_2}\left(\hatn\right)
     Y_{\left(J_3M_3\right)}\left(\hatn\right),
\end{equation}
for $\lambda_1,\lambda_2=0,\pm1$.
We use the completeness relation for the spherical basis,
\begin{equation}
     g^{ab}=\sum_{\lambda}\hat{\varepsilon}_{\lambda}^{a}
     (\hatn)\hat{\varepsilon}_{\lambda}^{b*}
     (\hatn)=\sum_{\lambda}\left(-1\right)^{\lambda}
     \hat{\varepsilon}_{\lambda}^{a}
     (\hatn)\hat{\varepsilon}_{-\lambda}^{b}
     (\hatn).
\end{equation}
The $g^{ab}$ on the far left does not really depend on $\hatn$,
but the decomposition can be done at each $\hat{n}$. Then, we find
\begin{eqnarray}
     \int d^2\hatn
     g^{ab}Y_{(J_1M_1)a}^{\lambda_1}(\hatn)
     Y_{(J_2M_2)b}^{\lambda_2}(\hatn)
     Y_{\left(J_3M_3\right)}(\hatn) & =&
     \sum_{\lambda}\left(-1\right)^{\lambda} \int
     d^2\hatn\hat{\varepsilon}_{\lambda}^{a}(\hatn)\hat{\varepsilon}_{-\lambda}^{b}
     (\hatn) Y_{(J_1M_1)a}^{\lambda_1}(\hatn)
     Y_{(J_2M_2)b}^{\lambda_2} (\hatn)
     Y_{\left(J_3M_3\right)}(\hatn) \nonumber  \\ 
      & =&
      \left(-1\right)^{\lambda_1}\delta_{\lambda_1,-\lambda_2}
      \int d^2\hatn {}_{-\lambda_1}Y_{(J_1M_1)}(\hatn)
      {}_{\lambda_1} Y_{(J_2M_2)}(\hatn)Y_{(J_3M_3)}(\hatn).
\label{eqn:threeY_vector_sYlm}
\end{eqnarray}
Here, we use Eq.~(52) of Ref.~\cite{Dai:2012} to relate the
vector spherical harmonics in the helicity basis to the
spin-weighted spherical harmonics \cite{NewmanPenrose}
\begin{equation}
     \hat{\varepsilon}_{\lambda'}^{a}(\hatn)
     Y_{(JM)a}^{\lambda}(\hatn) = {}_{-\lambda}Y_{(JM)}(\hatn)
     \delta_{\lambda\lambda'}.
\end{equation}

Since the sum of the three spins in the three spin-weighted
spherical harmonics in Eq.~(\ref{eqn:threeY_vector_sYlm}) is
zero, we use Eq.~(\ref{eqn:three_sYlm}) to calculate the angular
integral of three helicity-basis TAM waves to be,
\ba
     \int d^2\hatn g^{ab}
     Y_{(J_1M_1)a}^{\lambda_1}(\hatn)
     Y_{(J_2M_2)b}^{\lambda_2}(\hatn) Y_{\left(J_3M_3\right)}(\hatn)
     &=&\left(-1\right)^{\lambda_1}
     \delta_{\lambda_1,-\lambda_2} \mathcal{G}^{J_1 J_2
     J_3}_{M_1M_2M_3} \frac{\left<J_1 \lambda_1 J_2,
     -\lambda_1|J_30\right>}{\left<J_10J_20|J_30\right>}.
\ea
Due to the Kronecker delta, the only non-zero combinations are
\begin{eqnarray}
     \int d^2\hatn g^{ab}
     Y_{(J_1M_1)a}^{0}\left(\hatn\right)
     Y_{(J_2M_2)b}^{0}\left(\hatn\right)
     Y_{\left(J_3M_3\right)}\left(\hatn\right) & = &
     \mathcal{G}^{J_1 J_2 J_3}_{M_1M_2M_3} ,
     \nonumber \\
     \int d^2\hatn g^{ab}
     Y_{(J_1M_1)a}^{+1}\left(\hatn\right)
     Y_{(J_2M_2)b}^{-1}\left(\hatn\right)
     Y_{\left(J_3M_3\right)}\left(\hatn\right) & = &
     -\mathcal{G}^{J_1 J_2      J_3}_{M_1M_2M_3} \frac{\left<J_1 1 J_2,
     -1|J_30\right>}{\left<J_10J_20|J_30\right>},
    \nonumber \\
    \int d^2\hatn g^{ab}
    Y_{(J_1M_1)a}^{-1}\left(\hatn\right)
    Y_{(J_2M_2)b}^{+1}\left(\hatn\right)
    Y_{\left(J_3M_3\right)}\left(\hatn\right) & = &  - \mathcal{G}^{J_1 J_2
     J_3}_{M_1M_2M_3} \frac{\left<J_1, -1,J_2
     1|J_30\right>}{\left<J_10J_20|J_30\right>}.
\end{eqnarray}

\subsection{Two transverse tensor TAM-waves and one scalar TAM wave}
\label{app:3Ylms_LEB_tensor}

We now calculate the helicity-basis overlap integral,
\begin{equation}
     \int d^2\hatn
     I^{ab,cd}Y_{(J_1M_1)ab}^{\lambda_1}(\hatn)
     Y_{(J_2M_2)cd}^{\lambda_2}(\hatn)
     Y_{\left(J_3M_3\right)}(\hatn),
\end{equation}
for $\lambda_1,\lambda_2=0,\pm1,\pm2$.
With the completeness relation,
\begin{equation}
     I^{ab,cd}=\sum_{\lambda}\hat{\varepsilon}_{\lambda}^{ab}(\hatn)
     \hat{\varepsilon}_{\lambda}^{cd*}(\hatn) =
     \sum_{\lambda}\left(-1\right)^{\lambda}
     \hat{\varepsilon}_{\lambda}^{ab}(\hatn)
     \hat{\varepsilon}_{-\lambda}^{cd}(\hatn),
\end{equation}
and Eq.~(100) of Ref.~\cite{Dai:2012} (the relation between tensor
spherical harmonics and the spin-weight $s=2$ spherical harmonics),
\begin{equation}
     \hat{\varepsilon}_{\lambda'}^{ab}(\hatn)
     Y_{(JM)ab}^{\lambda}(\hatn) = {}_{-\lambda}Y_{(JM)}(\hatn)
     \delta_{\lambda\lambda'},
\end{equation}
the integration becomes
\begin{align}
     \int d^2\hatn\,
     I^{ab,cd}Y_{(J_1M_1)ab}^{\lambda_1}(\hatn)
     Y_{(J_2M_2)cd}^{\lambda_2}(\hatn)
     Y_{\left(J_3M_3\right)}(\hatn) & =
     \sum_{\lambda}\left(-1\right)^{\lambda} \int d^2\hat{n}
     \hat{\varepsilon}_{\lambda}^{ab}(\hatn)
     \hat{\varepsilon}_{-\lambda}^{cd}(\hatn)
     Y_{(J_1M_1)ab}^{\lambda_1}(\hatn)
     Y_{(J_2M_2)cd}^{\lambda_2}(\hatn)
     Y_{\left(J_3M_3\right)}(\hatn)\nonumber \\ 
     & =\left(-1\right)^{\lambda_1}
     \delta_{\lambda_1,-\lambda_2}\int
     d^2\hatn_{-\lambda_1}\,
     Y_{(J_1M_1)}(\hatn){}_{\lambda_1}
     Y_{(J_2M_2)}(\hatn) Y_{(J_3M_3)}(\hatn).
\end{align}
Then we find
\be
     \int d^2\hatn\, Y_{(J_1M_1)}^{\lambda_1,ab}(\hatn)
     Y_{(J_2M_2)ab}^{\lambda_2}(\hatn)Y_{(J_3M_3)}(\hatn)
   = (-1)^{\lambda_1}\delta_{\lambda_1,-\lambda_2}     
     \mathcal{G}^{J_1J_2J_3}_{M_1M_2M_3} \frac{
     \CG{J_1}{\lambda_1}{J_2}{,-\lambda_1}{J_3}{0}
     }{\CG{J_1}{0}{J_2}{0}{J_3}{0} }.
\ee

\section{A semi-classical interpretation}
\label{app:semi_classical}

In this Appendix, we briefly discuss a semi-classical picture
which allows us to interpret our results for the three-point
function involving two transverse vectors and one scalar in the
limit of large angular momentum.

In Sec.~\ref{sec:twoVoneS}, we presented three-point
correlations involving two transverse vectors and one scalar,
namely Eq.~(\ref{eqn:TAMbk_overlaps}), together with the overlap
integrals in Eqs.~(\ref{eqn:3ovl_BBs}), (\ref{eqn:3ovl_EEs}), and
(\ref{eqn:3ovl_BEs}).  Some re-arrangment of those results
enables us to rewrite them as
\begin{align}
\label{eq:3pt-EE-classical}
     \VEV{
     V_{(J_1M_1)}^{E,k_1}V_{(J_2M_2)}^{E,k_2}\phi_{(J_3M_3)}^{k_3}}
     &
     =-\left(4\pi\right)^{3}\left(-i\right)^{J_1+J_2+J_3}
     B_{vvs}\left(k_1,k_2,k_3\right)
     \mathcal{G}^{J_1J_2J_3}_{M_1M_2M_3} \nn\\
 & \times\int
 r^2dr\left[\frac{[1]+[2]-[3]}{2\sqrt{[1][2]}}j_{J_1,v}^{(E,E)}\left(k_1r\right)j_{J_2,v}^{(E,E)}\left(k_2r\right)+\sqrt{[1][2]}\frac{j_{J_1}\left(k_1r\right)}{k_1r}\frac{j_{J_2}\left(k_2r\right)}{k_2r}\right]j_{J_3}\left(k_3r\right),
\end{align}
\begin{align}
\langle
V_{(J_1M_1)}^{B,k_1}V_{(J_2M_2)}^{B,k_2}\phi_{(J_3M_3)}^{k_3}\rangle
&
=\left(4\pi\right)^{3}\left(-i\right)^{J_1+J_2+J_3}B_{vvs}\left(k_1,k_2,k_3\right)
 \frac{[1]+[2]-[3]}{2\sqrt{[1] [2]}}
     \mathcal{G}^{J_1J_2J_3}_{M_1M_2M_3} \nn\\
 & \times\int
 r^2drj_{J_1}\left(k_1r\right)j_{J_2}\left(k_2r\right)j_{J_3}\left(k_3r\right),
\end{align}
\begin{align}
\langle
V_{(J_1M_1)}^{B,k_1}V_{(J_2M_2)}^{E,k_2}\phi_{(J_3M_3)}^{k_3}\rangle
& =-\left(4\pi\right)^{3}\left(-i\right)^{J_1+J_2+J_3}
B_{vvs}\left(k_1,k_2,k_3\right)
\mathcal{G}^{J_1J_2J_3}_{M_1M_2M_3} \frac{ \CG{J_1}{0}{J_2}{0}{J_3-1,}{0}}{\CG{J_1}{0}{J_2}{0}{J_3}{0}} \sqrt{\frac{2J_3+1}{2J_3-1}}\nn\\
 &
 \times\frac{\sqrt{\left(-J_1+J_2+J_3\right)\left(J_1-J_2+J_3\right)\left(J_1+J_2-J_3+1\right)\left(J_1+J_2+J_3+1\right)}}{2\sqrt{[1][2]}}\nonumber
 \\
 & \times\int r^2drj_{J_1}\left(k_1r\right)j_{J_2,v}^{(E,E)}\left(k_2r\right)j_{J_3}\left(k_3r\right).
\end{align}
Here we have introduced the shorthand notation $[i] \equiv
J_i(J_i+1)$ for $i=1,2,3$. These three-point functions have now
been put into forms comparable to the three-point function,
Eq.~(\ref{eqn:scalarfinal}), for three scalars.
We have a bispectrum function in
terms of wave vectors with some suitable multiplying prefactor, a
Gaunt integral in accord with the Wigner-Eckart
theorem, and an integral of three radial profiles. For a
vector-vector-scalar correlation, we have an extra geometric
factor made of angular momenta $J_1$, $J_2$, and $J_3$. In the limit of
large angular momenta, i.e. $J_1,J_2,J_3 \gg 1$, these factors
are reduced to cosine and sine
\be
\frac{[1]+[2]-[3]}{2\sqrt{[1] [2]}}\approx\cos\theta_{12},\qquad
\frac{\sqrt{\left(-J_1+J_2+J_3\right)\left(J_1-J_2+J_3\right)\left(J_1+J_2-J_3+1\right)\left(J_1+J_2+J_3+1\right)}}{2\sqrt{[1][2]}}
\approx\sin\theta_{12}.
\ee
where $\theta_{12}$ is the angle between $J_1$ and $J_2$, in a triangle whose three sides have length $J_1$,$J_2$ and $J_3$, repectively. 

\begin{figure}
\centering
\includegraphics[scale=0.5]{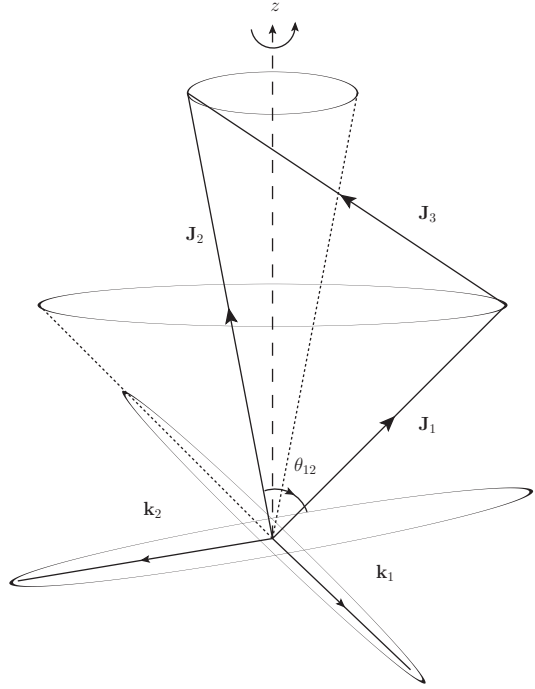}
\caption{Geometry of the semi-classical picture: three classical
     angular-momentum vectors
     $\mathbf{J}_1$, $\mathbf{J}_2$, and $\mathbf{J}_3$ form a triangle
     and precess about the chosen $z$-axis, associated with the
     three TAM waves of wave numbers $k_1$, $k_2$, and $k_3$
     respectively. The angle $\theta_{12}$ between
     $\mathbf{J}_1$ and $\mathbf{J}_2$, however, remains the
     same. The wave vectors $\mathbf{k}_1$ and $\mathbf{k}_2$, which
     are not definitive for TAM waves, are perpendicular to
     angular momenta $\mathbf{J}_1$ and $\mathbf{J}_2$, respectively.} 
\label{fig:classical_picture}
\end{figure}

These geometric factors, interpreted in a semi-classical way,
can be attributed to the vector structure of vector TAM
waves. First consider the case of two B-mode transverse
vectors.  The three-point function is proportional to the overlap,
\begin{align}
\int
d^{3}x\Psi_{(J_1M_1)}^{k_1,B,a}(\bfx)\Psi_{(J_2M_2)a}^{k_2,B}(\bfx)\Psi_{(J_3M_3)}^{k_3}(\bfx)
& =\int d^{3}x\left(\frac{-i\hat{\mathbf{L}}}{\sqrt{[1]}} \Psi_{(J_1M_1)}^{k_1}(\bfx)\right)\cdot\left(\frac{-i\hat{\mathbf{L}}}{\sqrt{[2]}}\Psi_{(J_2M_2)}^{k_2}(\bfx)\right)\Psi_{(J_3M_3)}(\bfx)\nonumber \\
 & \approx-\frac{1}{\sqrt{[1] [2]}} \int
 d^{3}x\mathbf{J}_1\Psi_{(J_1M_1)}\cdot\mathbf{J}_2\Psi_{(J_2M_2)}\Psi_{(J_3M_3)}\nonumber
 \\
 & =-\frac{1}{\sqrt{[1][2]}} \sqrt{[1]}\sqrt{[2]}\cos\theta_{12}\int d^{3}x\Psi_{(J_1M_1)}\Psi_{(J_2M_2)}\Psi_{(J_3M_3)}\nn\\
& =-\cos\theta_{12}\int d^{3}x\Psi_{(J_1M_1)}\Psi_{(J_2M_2)}\Psi_{(J_3M_3)},
\end{align}
where according to the construction in Ref.~\cite{Dai:2012} we
have used
$\Psi^{k,B}_{(JM)a}(\bfx)=K_a\Psi^{k}_{(JM)}(\bfx)/\sqrt{J(J+1)}$,
with $K_a=-i\hat{L}_a$ and $\hat{L}_a$ being the orbital
angular-momentum operator. In the second line, we associate two
classical angular-momentum vectors $\mathbf{J}_1,\mathbf{J}_2$,
of magnitude $J_1(J_1+1)$ and $J_2(J_2+1)$ respectively, with
the first two TAM waves. Note that $\mathbf{J}_1$ and $\mathbf{J}_2$
are not fixed vectors, since they precess about the
$z$ axis. But by angular-momentum conservation they always
differ by a third angular-momentum vector $\mathbf{J}_3$ of
magnitude $J_3(J_3+1)$ that is associated with the third TAM wave, which
also precesses about the $z$ axis, as shown in
Fig.~\ref{fig:classical_picture}. With this picture, the cosine
factor then arises naturally.

The case of one $B$ mode and one $E$ mode can be analyzed similarly,
\begin{align}
\int
d^{3}x\Psi_{(J_1M_1)}^{k_1,B,a}(\bfx)\Psi_{(J_2M_2)a}^{k_2,E}(\bfx)\Psi_{(J_3M_3)}^{k_3}(\bfx)
& =\int
d^{3}x\left(\frac{-i\hat{\mathbf{L}}}{\sqrt{[1]}}\Psi_{(J_1M_1)}^{k_1}(\bfx)\right)\cdot\left(\frac{\nabla\times\hat{\mathbf{L}}}{k_2\sqrt{[2]}}\Psi_{(J_2M_2)}^{k_2}(\bfx)\right)\Psi_{(J_3M_3)}(\bfx)\nonumber
\\ 
 & \approx\frac{1}{\sqrt{[1][2]}}\int d^{3}x\mathbf{J}_1\Psi_{(J_1M_1)}\cdot\left(\hat{\mathbf{k}_2}\times\mathbf{J}_2\Psi_{(J_2M_2)}\right)\Psi_{(J_3M_3)}\nonumber \\
 & =\left(\hat{\mathbf{J}}_1\cdot\left(\hat{\mathbf{k}}_2\times\mathbf{\hat{J}}_2\right)\right)\int d^{3}x\Psi_{(J_1M_1)}\Psi_{(J_2M_2)}\Psi_{(J_3M_3)}\nonumber \\
 & =-\left(\left(\hat{\mathbf{J}}_1\times\mathbf{\hat{J}}_2\right)\cdot\hat{\mathbf{k}}_2\right)\int d^{3}x\Psi_{(J_1M_1)}\Psi_{(J_2M_2)}\Psi_{(J_3M_3)},
\end{align}
where this time we have also used $\Psi^{k,B}_{(JM)a}(\bfx)=M_a\Psi^{k}_{(JM)}(\bfx)/\sqrt{J(J+1)}$, with $M_a=i\varepsilon_{abc}\nabla^b K^c/k$. We clearly see a factor $|\hat{\mathbf{J}}_1\times\hat{\mathbf{J}}_1|=\sin\theta_{12}$, which is the sine we find in the large-$J$ limit.

To close, we consider the case of two $E$-mode vectors
\begin{align}
\int
d^{3}x\Psi_{(J_1M_1)}^{k_1,E,a}(\bfx)\Psi_{(J_2M_2)a}^{k_2,E}(\bfx)\Psi_{(J_3M_3)}^{k_3}(\bfx)
& =\int
d^{3}x\left(\frac{\nabla\times\hat{\mathbf{L}}}{k_1\sqrt{[1]}}
\Psi_{(J_1M_1)}^{k_1}(\bfx)\right)\cdot\left(\frac{\nabla\times\hat{\mathbf{L}}}{k_2\sqrt{[2]}}\Psi_{(J_2M_2)}^{k_2}(\bfx)\right)\Psi_{(J_3M_3)}\left(\bfx
\right)\nonumber \\
 & \approx-\frac{1}{\sqrt{[1][2]}}\int d^{3}x\left(\hat{\mathbf{k}}_1\times\mathbf{J}_1\right)\Psi_{(J_1M_1)}\cdot\left(\hat{\mathbf{k}_2}\times\mathbf{J}_2\Psi_{(J_2M_2)}\right)\Psi_{(J_3M_3)}\nonumber \\
 & \approx\left(\hat{\mathbf{k}}_1\times\mathbf{\hat{J}}_1\right)\cdot\left(\hat{\mathbf{k}}_2\times\mathbf{\hat{J}}_2\right)\int d^{3}x\Psi_{(J_1M_1)}\Psi_{(J_2M_2)}\Psi_{(J_3M_3)}\nonumber \\
 & =\left[\left(\hat{\mathbf{k}}_1\cdot\hat{\mathbf{k}}_2\right)\left(\hat{\mathbf{J}}_1\cdot\hat{\mathbf{J}}_2\right)-\left(\hat{\mathbf{k}}_1\cdot\hat{\mathbf{J}}_2\right)\left(\hat{\mathbf{J}}_1\cdot\hat{\mathbf{k}}_2\right)\right]\int d^{3}x\Psi_{(J_1M_1)}\Psi_{(J_2M_2)}\Psi_{(J_3M_3)}.
\end{align}
We then recognize a term proportional to
$\hat{\mathbf{J}}_1\cdot\hat{\mathbf{J}}_2$,
i.e. $\cos\theta_{12}$, and a second term which has no such
factor. It closely resembles the structure in the exact result
Eq.~(\ref{eq:3pt-EE-classical}).

So far, we have restricted our discussion to the three-point
function involving two transverse vectors and one
scalar. Besides, the semi-classical picture we have proposed is
still insufficient to quantitatively pin down the correct forms
of the radial integral and the $J_1,J_2,J_3$
dependence. However, we may gain insight from this picture
for other three-point functions, such as the ones involving
transverse-tensor field.

\end{document}